# Was Venus Ever Habitable? Constraints from a Coupled Interior–Atmosphere–Redox Evolution Model

Joshua Krissansen-Totton[1,3], Jonathan J. Fortney[1] , and Francis Nimmo[2]
[1] Department of Astronomy and Astrophysics, University of California, Santa Cruz, CA 95064, USA; jkt@ucsc.edu
[2] Department of Earth and Planetary Sciences, University of California, Santa Cruz, CA 95064, USA
*Received 2021 March 4; revised 2021 August 12; accepted 2021 September 7; published 2021 October 29*

## Abstract

Venus's past climate evolution is uncertain. General circulation model simulations permit a habitable climate as late as ∼0.7 Ga, and there is suggestive—albeit inconclusive—evidence for previous liquid water from surface geomorphology and mineralogy. However, it is unclear whether a habitable past can be reconciled with Venus's inferred atmospheric evolution. In particular, the lack of leftover atmospheric oxygen argues against recent water loss. Here, we apply a fully coupled model of Venus's atmospheric–interior–climate evolution from post-accretion magma ocean to present. The model self-consistently tracks C-, H-, and O-bearing volatiles and surface climate through the entirety of Venus's history. Atmospheric escape, mantle convection, melt production, outgassing, deep water cycling, and carbon cycling are explicitly coupled to climate and redox evolution. Plate tectonic and stagnant lid histories are considered. Using this coupled model, we conclude that both a habitable Venusian past and one where Venus never possessed liquid surface water can be reconciled with known constraints. Specifically, either scenario can reproduce bulk atmospheric composition, inferred surface heat flow, and observed $^{40}Ar$ and $^{4}He$. Moreover, the model suggests that Venus could have been habitable with a ∼100 m global ocean as late as 1 Ga, without violating any known constraints. In fact, if diffusion-limited water loss is throttled by a cool, $CO_2$-dominated upper atmosphere, then a habitable past is tentatively favored by our model. This escape throttling makes it difficult to simultaneously recover negligible water vapor and ∼90 bar $CO_2$ in the modern atmosphere without temporarily sequestering carbon in the interior via silicate weathering to enhance H escape.

*Unified Astronomy Thesaurus concepts:* Venus (1763); Solar system terrestrial planets (797); Planetary atmospheres (1244); Planetary climates (2184); Atmospheric composition (2120); Planetary interior (1248); Tectonics (2175); Volcanoes (1780); Habitable planets (695)

*Supporting material:* animation

## 1. Introduction

Despite decades of study, Venus's past remains enigmatic. Models of planetary formation suggest that early Venus ought to have been endowed with a water inventory comparable to that of Earth (Raymond et al. 2006, 2007), and elevated D/H ratios in the modern atmosphere are indicative of past water loss (Donahue et al. 1982). But whether Venus's surface was ever habitable is a subject of ongoing debate. Understanding past conditions on Venus may enable more robust interpretations of terrestrial exoplanet observations (Kane et al. 2019; Lustig-Yaeger et al. 2019b). Moreover, purported detections of phosphine in Venus's atmosphere and its possible biological origin (Bains et al. 2021; Greaves et al. 2021), while controversial (Encrenaz et al. 2020; Snellen et al. 2020; Akins et al. 2021; Villanueva et al. 2021), similarly highlight the need to better understand Venusian history, as does recent reanalysis of Pioneer Mass Spectrometer data that are suggestive of previously unknown chemical disequilibria in Venus's middle atmosphere (Mogul et al. 2021).

Models of Venusian climate evolution are conflicted on whether Venus experienced transient surface habitability early in its history. Kasting (1988) used a radiative-convective climate model to show that the insolation received by early Venus lies close to the runaway greenhouse threshold. Consequently, whether early Venus possessed liquid surface water or has always been in a runaway greenhouse state depends on cloud feedbacks. This result has been borne out by models coupling surface climate to magma ocean evolution, which show that Venus sits on the threshold between long-lived magma ocean (never habitable) and a brief initial magma ocean followed by ocean condensation (Hamano et al. 2013). Cloud-free, 1D radiative-convective models tend to predict a low runaway greenhouse threshold, thereby suggesting that Venus has always been in a runaway greenhouse state since formation (Goldblatt et al. 2013; Ramirez et al. 2014). However, 3D GCM models have revealed that cloud feedbacks, specifically high reflectivity caused by slow rotation and enhanced dayside cloudiness, could have maintained temperate conditions on Venus until 0.7 Ga (Way et al. 2016). Subsequent GCM modeling showed that these cloud feedbacks could have even maintained habitable surface conditions until today (Way & Del Genio 2020). In this view, the transition from habitable to uninhabitable surface conditions was not necessarily caused by insolation increases triggering a runaway greenhouse, but instead by massive $CO_2$ outgassing coincident with resurfacing around 0.7 Ga (Way & Del Genio 2020). In addition to insolation changes and interior evolution forcings, orbital and tidal interactions may have also contributed to Venus's atmospheric evolution (Innanen et al. 1998; Barnes et al. 2013; Kane et al. 2020).

---

[3] NASA Sagan Fellow.

The plausibility of past habitable climates depends on whether these scenarios can be reconciled with Venusian atmospheric evolution as constrained by observed composition and the history of atmospheric escape. Crucially, Venus's modern atmosphere contains virtually no molecular oxygen, with a <3 ppm upper limit above ~60 km (Mills 1999). This lack of oxygen must be reconciled with atmospheric water loss since the escape of large amounts of hydrogen would seemingly predict significant "leftover" oxygen accumulation. Kasting & Pollack (1983) established that the hydrodynamic loss of H is a viable mechanism for extensive early atmospheric loss and that this may have contributed to observed D/H fractionation. Zahnle & Kasting (1986) modeled this hydrodynamic loss in more detail and concluded that a large amount of oxygen could have been dragged off to space given the high X-ray and ultraviolet (XUV) of the early Sun. Subsequent modeling of hydrodynamic H and O loss by Chassefière (1996) downgraded these O drag fluxes and showed that, in the absence of other substantial oxygen sinks, the initial water inventory could not have exceeded 0.45 Earth oceans without avoiding substantial atmospheric oxygen accumulation (the modern Venusian atmosphere has essentially no $O_2$).

In light of this constraint, Gillmann et al. (2009) proposed a self-consistent scenario for early hydrodynamic H loss. Starting with an initial inventory of <5 terrestrial oceans, oxygen produced by early H loss may have been sequestered in an early magma ocean, thereby avoiding later buildup. Vigorous H escape during this initial magma ocean phase can also reproduce observed Ar and Ne isotope ratios. Loss of water would eventually trigger the solidification of the magma ocean, leaving behind small amounts of H and O, which could plausibly be lost to space and consumed by surface sinks, respectively (Gillmann et al. 2009). Such a scenario precludes large surface oceans following magma ocean solidification but is agnostic on subsequent surface habitability with small water reservoirs (Chassefière et al. 2012). Broadly speaking, however, a Venus that has been in a runaway greenhouse state since formation is easier to reconcile with the current lack of atmospheric oxygen. This is because recent water loss means less solar XUV flux to drive O drag, and thus larger interior sinks are required to mop up atmospheric oxygen (e.g., Lammer et al. 2018).

Isotopes and trace gases have also been used to constrain Venus's atmospheric evolution. While observed D/H enrichment is suggestive of water loss (Donahue et al. 1982), this is a highly degenerate constraint. Vigorous hydrodynamic escape or impact-induced atmospheric loss may not yield any isotopic fractionation, and so observed D/H ratios provide a lower limit on water loss (Kasting & Pollack 1983; Zahnle et al. 1988). Moreover, it may be possible to explain the observed D/H ratio as a steady-state balance between recent water loss and resupply as opposed to the remnant from an initial water inventory (Grinspoon 1987), although the former requires a contemporary water source more D/H-rich than comets (Grinspoon 1993). Uncertainties in D/H fractionation by nonthermal escape processes and variation in D/H with altitude (Marcq et al. 2018) also complicate water loss interpretations. Observed isotopic ratios of $^{36}Ar/^{38}Ar$ and $^{20}Ne/^{22}Ne$ seem to favor a rapid accretion (Lammer et al. 2020), but these constraints are consistent with a number of early escape scenarios and are largely agnostic on subsequent habitability.

Venusian geomorphology and mineralogy may also provide insights into climate evolution. The scarcity of craters and apparent lack of crater modification revealed by radar imagery imply an average surface age <1 Gyr and have been conventionally interpreted as reflecting a recent resurfacing event (Schaber et al. 1992; Strom et al. 1994), although equilibrium resurfacing cannot be excluded (Bjonnes et al. 2012). Indeed, ~80% of craters have radar-dark floors, which have been interpreted as postimpact lava flows (Herrick & Rumpf 2011), which suggests that some combination of continuous, local resurfacing and regional lava flows could explain the crater distribution without recourse to catastrophic resurfacing (O'Rourke et al. 2014). Resurfacing, whether by catastrophic event or gradual crustal turnover, means that most geologic records of Venus's early history have been erased. However, some surface regions, such as the highly deformed tessera terrain, are conjectured to predate resurfacing (Hansen & López 2010). This is of interest for past climate because some tessera regions also have low-emissivity surfaces attributed to felsic crust (Gilmore et al. 2015, 2017). Venera 13 elemental abundance measurements are also suggestive of elevated $K_2O$, felsic crust (McKenzie et al. 1992). The presence of felsic crust is significant because it may indicate past surface water, although felsic crust formation can also be formed by differentiation of anhydrous basaltic melts (Gilmore et al. 2015). Moreover, detections of felsic crust are not definitive (Wroblewski et al. 2019). The tessera terrain show signs of erosion (Byrne et al. 2021), potentially due to past fluvial processes (Khawja et al. 2020). The observed erosion features could also be aeolian (Byrne et al. 2021), however, and so neither mineralogical nor geomorphological observations necessarily require a past hydrosphere.

Coupled atmospheric-interior models constrained by modern atmospheric composition and surface conditions may also be used to provide insight into Venus's past evolution. Early conceptual models highlighted possible feedbacks between surface climate and melt production (Phillips et al. 2001). Driscoll & Bercovici (2013) used a carbon cycle box model with a gray atmosphere to illustrate the divergent climate histories of Venus and Earth. Sophisticated 2D and 3D mantle convection models have also been applied to elucidate feedbacks between tectonics, resurfacing, outgassing, and atmospheric evolution (Noack et al. 2012; Gillmann & Tackley 2014). Gillmann et al. (2020) added late accretion impact scenarios to a coupled atmosphere-interior evolution model to determine how the modern atmosphere could be recovered. They found that late accretion materials must have been relatively volatile poor and that the total mass delivered during late accretion is limited by the inefficient oxygen loss during the later stages of Venus's evolution.

None of these coupled atmospheric-interior models explicitly explore the possibility of previous liquid surface water and consistency with atmospheric redox evolution, however. Here, we apply a coupled model of Venus's evolution from post-accretion magma ocean to the present, with the goal of determining whether a habitable past is consistent with all known atmospheric and surface constraints. Radiative-convective climate, atmospheric escape, parameterized mantle convection, and melt production/outgassing models are coupled to track fluxes of H-, C-, and O-bearing species between the atmosphere and the interior and to constrain the evolution of any surface liquid water inventory. The possibility of subduction returning carbonates, hydrous minerals, or oxidized crust to the mantle during temperate periods early in Venus's evolution is included. Moreover, we explicitly model the redox evolution of the mantle and atmosphere to evaluate whether habitable and nonhabitable scenarios are consistent with the current lack of oxygen in the modern atmosphere.

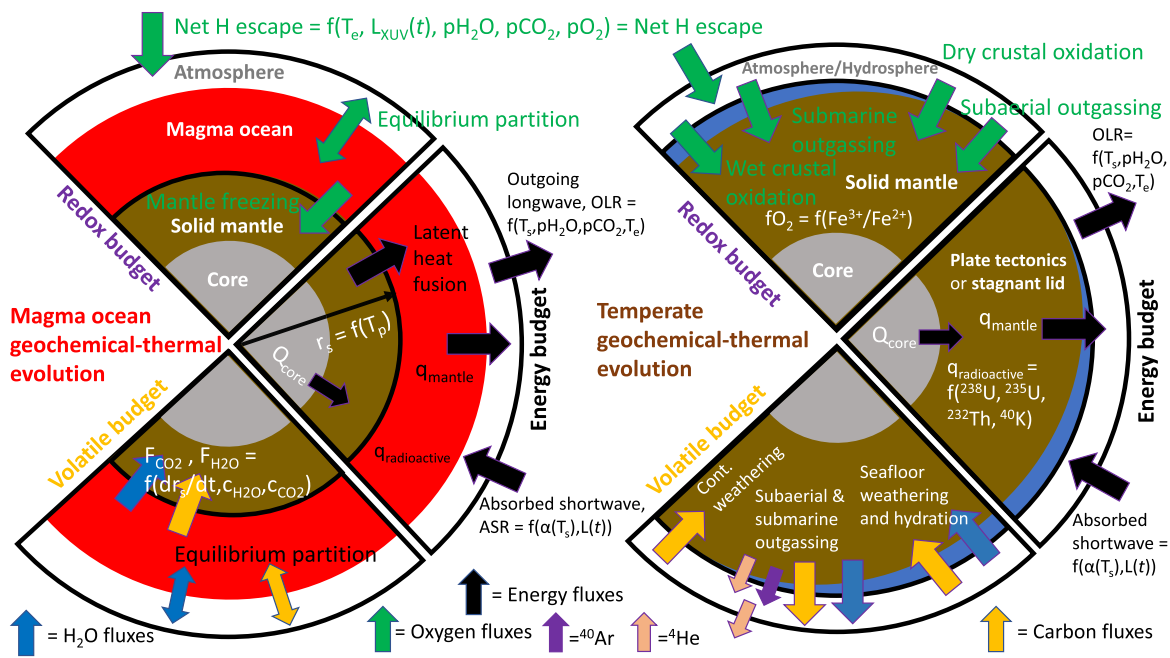

**Figure 1.** Schematic of the PACMAN geochemical evolution model applied to Venus. The redox budget, thermal-climate evolution, and volatile budget of Venus are modeled from initial magma ocean (left) through to temperate geochemical cycling (right). Oxygen fluxes are shown by green arrows, energy fluxes by black arrows, carbon fluxes by orange arrows, and water fluxes by blue arrows. Note that the net loss of H to space effectively adds oxygen to the atmosphere. During the magma ocean phase, the radius of solidification, $r_s$, begins at the core–mantle boundary and moves toward the surface as internal heat is dissipated. The rate at which this occurs is controlled by radiogenic heat production, $q_r$, convective heat flow from the mantle to the surface, $q_{mantle}$, and heat flow from the core, $Q_{core}$. This internal heat flow balances the difference between outgoing longwave radiation (OLR) and incoming absorbed shortwave radiation (ASR). The oxygen fugacity of the mantle, $fO_2$, and the water and carbon content of mantle and surface reservoirs are tracked throughout. Both plate tectonics and stagnant lid modes are permitted, and $^{40}$Ar (purple arrow) and $^4$He (beige arrows) atmospheric accumulation is explicitly modeled. This schematic is an adapted and updated version from Krissansen-Totton et al. (2021).

## 2. Methods

In this study, the Planetary Atmosphere, Crust, and MANtle (PACMAN) geochemical evolution model is used to study Venus.[4] The model is adapted from Krissansen-Totton et al. (2021) and is qualitatively summarized in Figure 1. The reader is referred to this paper for the full description of the PACMAN model. Here we summarize the key features necessary for interpreting results. Modifications to the model for application to Venus are also described below and in Appendix A.

Planetary evolution is divided into an initial magma ocean phase and a subsequent solid-mantle phase. Post-core formation Venus is initialized with a fully molten mantle and some endowment of volatiles, a radionuclide inventory ($^{238}$U, $^{235}$U, $^{232}$Th, $^{40}$K), and an initial mantle oxygen fugacity (Figure 1, left). The magma ocean freezes from the core upward, and H, C, and O are partitioned between dissolved melt phases, crystalline phases, and the atmosphere by assuming chemical equilibrium. Heat flow from the interior is calculated using 1D convective parameterization, with temperature-dependent magma ocean viscosity.

Planetary oxidation may occur from the loss of hydrogen to space (less oxygen drag); free oxygen produced via H escape is dissolved in the melt and may be transferred to the solid mantle as the magma ocean solidifies (Schaefer et al. 2016). We parameterized atmospheric escape as either diffusion limited or XUV limited, depending on the composition of the mesosphere and the stellar XUV flux. During XUV-driven escape of a steam-dominated atmosphere, the hydrodynamic escape of H may drag along O (and even $CO_2$) following Odert et al. (2018). In contrast, if the upper atmosphere is mostly dry, then the escape of H will be limited by the diffusion of hydrogen through the background noncondensable atmosphere, and neither O nor $CO_2$ can escape. While only thermal escape is considered in the nominal model, we also perform sensitivity tests that include the nonthermal escape of oxygen ions over Venus's history (Kulikov et al. 2006; Persson et al. 2020).

A radiative-convective climate model is used to self-consistently calculate surface temperature, outgoing longwave radiation (OLR), absorbed shortwave radiation (ASR), the water vapor profile, and surface liquid water inventory (if any) during both the magma ocean phase and subsequent temperate evolution. OLR is a function of the surface $H_2O$, $CO_2$, and background $N_2$ inventories and is calculated using the publicly available correlated-$k$ radiative transfer code of Marcq et al. (2017). This code uses DISORT (Stamnes et al. 1988) with four-stream longwave radiative transfer. Correlated-$k$ coefficients are calculated from the high-resolution molecular absorption spectra computed with *kspectrum* (Eymet et al. 2016), $H_2O$–$H_2O$ continuum absorption is taken from Clough et al. (2005), and $CO_2$–$CO_2$ continuum absorption from fits to Venus observations (Bézard et al. 2011). $H_2O$–$CO_2$ continuum opacity is not considered and is likely negligible compared to $H_2O$–$H_2O$ and $CO_2$–$CO_2$ continuum absorption (Ma & Tipping 1992). The runaway greenhouse limit calculated using the code of Marcq et al. (2017) closely agrees with line-by-line calculations in Goldblatt et al. (2013). To obtain OLR in the presence of condensable water vapor, a dry adiabat to moist adiabat to isothermal atmospheric structure is assumed (Kasting 1988). While opacity sources are accurately quantified, atmospheric chemistry and cloud feedbacks under high pressure and/or steam-dominated atmospheres are challenging to model (e.g., Goldblatt et al. 2013; Marcq et al. 2017; Bierson & Zhang 2020; Boukrouche et al. 2021), especially for transition states between temperate and runaway greenhouse states. The atmospheric model outputs in this study should thus be taken as best estimates rather than rigorous climate calculations. In nominal calculations, the temperature of the isothermal upper atmosphere is randomly sampled from 180 to 220 K (see below), but in subsequent sensitivity tests upper atmosphere temperature is allowed to vary as a function of $pCO_2$ (Wordsworth & Pierrehumbert 2013).

---

[4] Code available on GitHub: https://github.com/joshuakt/Venus-evolution.

Loss of water to space and dissipation of internal heat from accretion and radionuclides eventually cause Venus's surface temperature to drop below the solidus. When this occurs, the magma ocean phase is complete, and the mantle transitions to solid-state mantle convection with temperate geochemical cycling (Figure 1, right). During solid-state evolution, the only source of oxygen remains atmospheric H escape. However, there are now numerous crustal sinks for oxygen, including both subaerial and submarine outgassing of reduced species (e.g., $H_2$, CO, $CH_4$), water-rock reactions that generate $H_2$, and dry crustal oxidation. The sizes of these oxygen sinks are self-consistently calculated from the planetary interior evolution and mantle volatile content: outgassing fluxes are calculated using the melt-gas equilibrium outgassing model of Wogan et al. (2020). Sulfur species are ignored in our nominal model because, for Earth-like bulk abundances, outgassing of S-bearing gases only modestly increases total oxygen sinks (Krissansen-Totton et al. 2021). Outgassing fluxes depend on mantle oxygen fugacity, degassing overburden pressure, the volatile content of the mantle, specifically $H_2O$ and $CO_2$ content, the rate at which melt is produced, and the tectonic regime (see below). Dry and wet crustal sinks for oxygen similarly depend on crustal production rates.

To estimate crustal production rates, both plate tectonics (Krissansen-Totton et al. 2021) and stagnant lid regimes (Foley & Smye 2018) are explicitly modeled. For full generality, we assume a transition from plate tectonics to stagnant lid anytime from 0.05 to 4 Gyr after magma ocean solidification (randomly sampled). This covers both end-member cases for Venus's tectonic evolution and ensures a stagnant lid regime for at least the last ∼0.5 Gyr. During plate tectonics, outgassing fluxes depend on the product of crustal depth, spreading rate, and total ridge length. During stagnant lid, outgassing fluxes are controlled by the upwelling flux of melt beneath the rigid lid, as determined by a characteristic convection velocity and convection cell length. The evolution of lithospheric and crustal thickness is modeled following Foley & Smye (2018). The thermal evolution of the mantle during solid-state evolution is calculated as follows: heat production from radionuclides and heat flow from the metallic core is balanced by convective and advective transport to the surface. Core heat flow is imposed and is assumed to be at the upper limit of what is consistent with the absence of a modern dynamo (Nimmo 2002; Driscoll & Bercovici 2014). Appendix A provides further details on internal evolution and tectonic parameterizations.

Silicate weathering (Krissansen-Totton et al. 2018) and the deep hydrological cycle (Schaefer & Sasselov 2015) are also explicitly modeled because climate and surface volatile inventories control crustal oxygen sinks and atmospheric escape processes. Carbon is added to the atmosphere via magmatic outgassing (described above) and—if surface liquid water exists—is returned via continental and seafloor weathering, whose relative contributions depend on climate and the total surface water inventory.

### 2.1. Treatment of Volatiles

The model tracks C-, H-, and O-bearing species, as well as $Fe^{3+}/Fe^{2+}$ speciation in the interior. Specifically, surface and mantle inventories of $CO_2$, $H_2O$, and $O_2$ are explicitly calculated. While our outgassing model self-consistently calculates outgassing fluxes of $CO_2$, $H_2O$, CO, $H_2$, and $CH_4$ from melt properties (Wogan et al. 2020) and $H_2$ fluxes from serpentinization, all outgassed reductants (CO, $H_2$, $CH_4$) are assumed to instantaneously deplete atmospheric oxygen. In the absence of atmospheric $O_2$, all H from H-bearing reductants is assumed to be rapidly lost to space, therefore not oxidizing the planet; H escape via outgassed $H_2$ and $CH_4$ will be rapid because, unlike $H_2O$, these species are not cold trapped. The limitations of our climate module and absence of a photochemical model mean that we cannot simulate the time evolution of more reducing bulk atmospheres (i.e., CO–$H_2$ dominated). An oxidized atmosphere is a common assumption in Venus evolution models (e.g., Driscoll & Bercovici 2013; Gillmann & Tackley 2014; Gillmann et al. 2020), but the possible limitations of this assumption are explored in Section 4, and sensitivity tests are presented. During melt production, volatiles are partitioned between solid and melt phases using a constant partition coefficient for water ($k_{H_2O} = 0.01$) and redox-dependent graphite-saturated partitioning for carbon (see Text G in Krissansen-Totton et al. 2021, which follows Ortenzi et al. 2020). Nitrogen fluxes are not modeled, and we instead assumed a constant $N_2$ background partial pressure in all model runs: 1 bar is assumed in the nominal model, but sensitivity tests for 0.5–3 bar $N_2$ partial pressures are also presented.

### 2.2. Monte Carlo Calculations and Unknown Parameters

There are many unknown parameters and initial conditions in the model, including temperature-dependent mantle viscosities, efficiencies of XUV-driven escape, uncertain early Sun XUV fluxes, carbon cycle feedbacks, deep hydrological cycle dependencies, tectonic regime transition time, and albedo parameterizations. We adopted a Monte Carlo approach, where we ran the model thousands of times uniformly sampling all 24 unknown parameters. Parameter ranges for key variables are shown in Table A1. Unknown parameters that are particularly important for Venus's evolution include the dry crustal oxidation efficiency, $f_{dry-oxid}$, which is the fraction of $Fe^{2+}$ in newly produced crust that is oxidized to $Fe^{3+}$ in the presence of an oxidizing atmosphere via nonaqueous reactions. This parameter is sampled uniformly in log space from $10^{-4}$ to 10% and includes a range of physical processes, including the diffusion of oxygen into extrusive lava flows (Gillmann et al. [2009]), direct oxidation of small grain erosion products (Arvidson et al. [1992]), and other gas–solid redox reactions (Zolotov [2019]), including aerosol oxidation following explosive volcanism (Warren & Kite [2021]). We endow Venus with anywhere from 2% to 200% Earth oceans by mass, and ∼40–600 bar $CO_2$ (extending these ranges to larger initial inventories does not recover modern Venus constraints). Initial free O ($2 \times 10^{21}$–$6 \times 10^{21}$ kg) is chosen to recover an Earth-like modern mantle oxygen fugacity, but sensitivity tests are considered with more reducing mantles. Parameter ranges for mantle viscosity span three orders of magnitude and were chosen to broadly represent silicate mantles of Earth-sized planets (Krissansen-Totton et al. [2021]), and as noted above, the transition time from plate tectonics to stagnant lid is sampled from ∼ 4.5 to 0.5 Ga for full generality (Table [A1]). Parameter ranges governing carbon cycle and deep hydrological cycle feedbacks were derived from Earth values (Krissansen-Totton et al. [2021]).

Another important parameter is planetary albedo. To calculate ASR over a wide range of temperatures, we adopted the albedo parameterization described in Pluriel et al. ([2019]), which defines a temperate state albedo and a runaway greenhouse state albedo. Rather than explicitly model complex cloud feedbacks, the albedo of the temperate state is a free parameter ranging from 0.2 to 0.7. Although we are not explicitly modeling cloud feedbacks, this agnostic approach allows us to test to what extent cloud feedbacks maintaining habitability are consistent with Venus's thermal, geochemical, and atmospheric evolution.

In addition to tracking C-, H-, and O-bearing species, the Venus model calculates the accumulation of atmospheric $^{40}Ar$ and $^4He$. $^{40}Ar$ is a decay product of $^{40}K$ and is released into the atmosphere when K- and Ar-bearing mantle rock is melted and brought to the surface. Since $^{40}Ar$ is not removed from the atmosphere once degassed, the total inventory is a measure of cumulative magmatic activity over a planet's history (Pollack & Black [1982]; Kaula [1999]; O'Rourke & Korenaga [2015]). Similarly, $^4He$ is a decay product of uranium and thorium that is released during outgassing, but unlike $^{40}Ar$, $^4He$ is lost to space on $10^8$ yr timescales (Krasnopolsky & Gladstone [2005]). The atmospheric inventory of $^4He$ is therefore a measure of more recent magmatic activity (Chassefiere et al. [1986]; Namiki & Solomon [1998]; Fedorov et al. [2011]). Including $^{40}Ar$ and $^4He$ in our model provides an additional way for testing for self-consistency against the observed Venusian atmosphere and enables us to determine whether a habitable or nonhabitable Venusian past better predicts the modern atmosphere.

## 3. Results

The coupled model was run 10,000 times randomly sampling all 24 unknown parameters, and the time evolution of each model run spans from post-accretionary magma ocean to the present day. We focus on only those model runs that could reproduce modern Venus atmospheric composition (Arney et al. [2014]), as defined by 40–200 mbar $pCO_2$, $< \sim 0.2$ mbar $pO_2$ ($< 10^{15}$ kg), and $< \sim 3$ mbar $pH_2O$ ($< 2 \times 10^{16}$ kg). We do not require strict consistency with observed $O_2$ upper limits (Mills [1999]) because we omit photochemistry. In any case, modifying these acceptance thresholds does not change any of our qualitative conclusions.

Of the 10,000 model runs, about 10% successfully recover modern Venus as defined by these atmospheric criteria.

Moreover, these Venus reproductions fall neatly into two categories: those where Venus was never habitable, and those where Venus experienced an epoch of surface habitability before returning to runaway greenhouse.

Figure [2] shows all model runs that successfully reproduce the modern Venusian atmosphere and where Venus was never habitable (i.e., no surface liquid water). Figure [2](a) shows the thermal evolution of the mantle and the surface from magma ocean to today, where mantle potential temperature is the temperature a parcel of mantle would have if it ascended to the surface adiabatically. Figure [2](b) shows the evolution of the magma ocean radius of solidification. Consistent with previous studies (Hamano et al. [2013]), we find that the duration of the magma ocean depends on planetary albedo and initial water abundance. Figure [2](c) shows the coupled evolution of atmospheric carbon dioxide, water, and oxygen. Note the catastrophic degassing of water during magma ocean solidification and the subsequent gradual loss of water to space. Water dissociation and the subsequent escape of hydrogen leave molecular oxygen, which is then consumed by crustal sinks. Figure [2](d) confirms that liquid water never condenses on the surface in any model runs. Figure [2](e) shows the planetary energy budget, including ASR, OLR, heat flow from the interior ($q_m$), and the runaway greenhouse limit for reference. The onset/end of a runaway greenhouse state occurs when the absorbed shortwave crosses the runaway greenhouse threshold. The model successfully recovers estimates of the modern heat flow, around 20 mW m$^{-2}$ (Nimmo & McKenzie [1998]).

Figures [2](g) and (j) show the evolution of melt production and crustal depth, respectively. By definition, both melt production and crustal depth assume positive values only after magma ocean solidification is complete. The modern crustal depth is broadly consistent with estimates from the literature ranging from 20 to 60 km (Smrekar [1994]; Simons et al. [1997]; Nimmo & McKenzie [1998]), although our crustal thickness may overestimate the modern crust since we are ignoring foundering below the basalt–eclogite transition (O'Rourke & Korenaga [2012]). The abrupt transition in crustal thickness marks the change from a plate tectonics regime to stagnant lid regime.

Figure [2](h) shows the evolution of mantle redox state relative to the quartz-fayalite-magnetite (QFM) buffer, and Figure [2](i) shows the redox budget of the atmosphere. The net flux in Figure [2](i) dictates the evolution of atmospheric oxygen in Figure [2](c). Hydrogen escape (minus oxygen drag) is a source of oxygen, whereas the oxidation of newly produced crust ("dry crustal") is a sink. Note that in this scenario there is no oxygen consumption via water–rock reactions because water never condenses on the surface. Additionally, both water and carbon outgassing are negligible (Figure [2](f)) because, even though melt production is high (Figure [2](g)), the mantle is left relatively desiccated after the magma ocean phase. This, combined with the high pressures of a thick $CO_2$-dominated atmosphere, means that what few mantle volatiles remain tend to partition into the melt and do not degas into the atmosphere (Figure [2](f)). Volatile quantities in the solid mantle as a fraction of total mantle and surface volatile reservoirs are plotted in Figure [2](l). The sensitivity of our results to retention of volatiles during magma ocean solidification is explored below, and we find that greater retention of volatiles in the mantle does not qualitatively change this scenario. Finally, we note that these model runs also reproduce the observed $^{40}Ar$ abundance in the atmosphere and make $^4He$ abundance

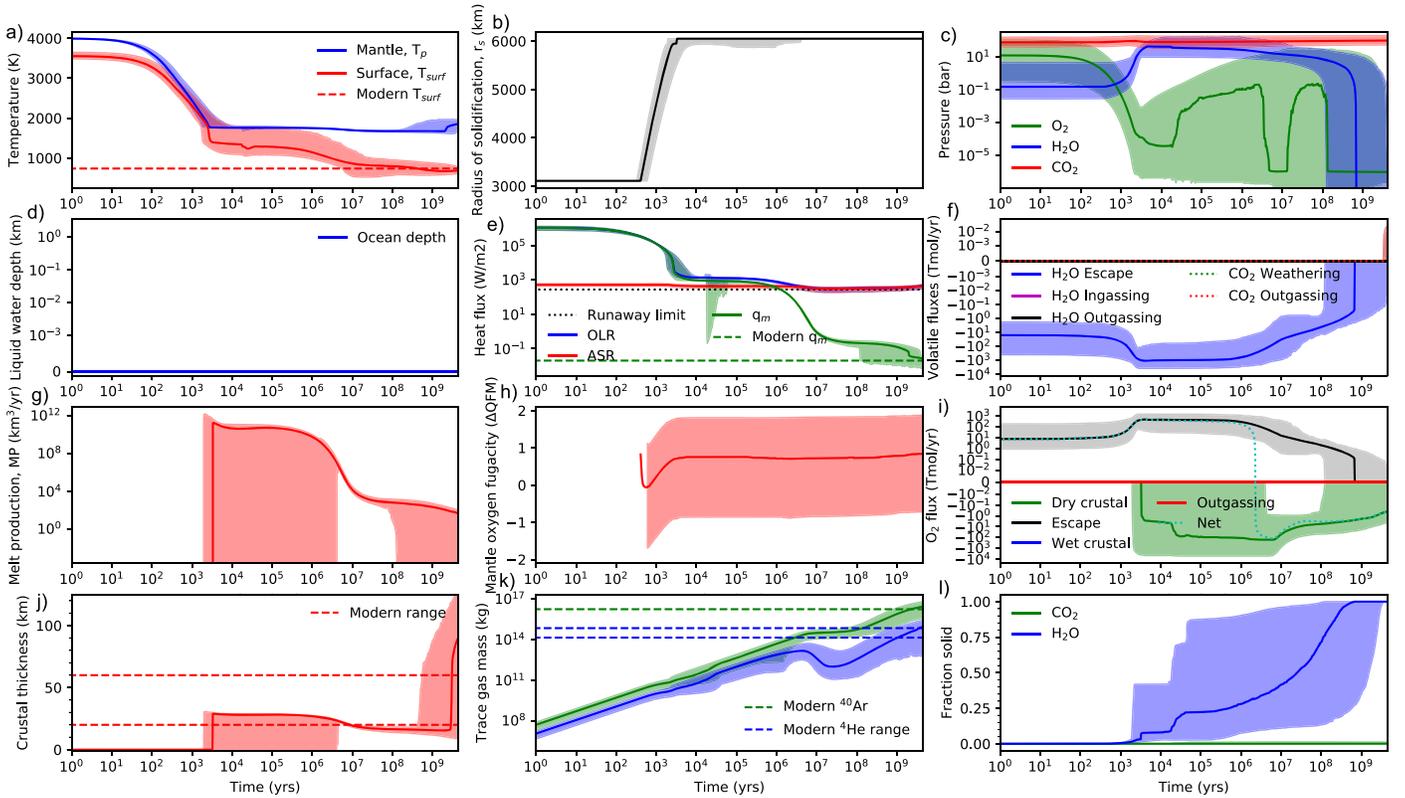

**Figure 2.** Model runs that successfully reproduce modern Venus's bulk atmosphere and where past surface conditions were never habitable. The lines are median values, and shaded regions denote 95% confidence intervals. The time evolution spans from post-accretionary magma ocean to the present day (4.5 × 10⁹ yr). Panel (a) shows the evolution of mantle potential temperature (blue) and surface temperature (red) over the last 4.5 Gyr. The initial magma ocean (panel (b)) persists from anywhere between a few thousand years to millions of years, depending on initial volatile endowments and planetary albedo. Water catastrophically degasses from the magma ocean and is subsequently lost to space via hydrodynamic escape (panels (c) and (f)). In these never-habitable model runs, the absorbed shortwave radiation never drops below the runaway greenhouse limit (panel (e)), and so liquid water never condenses on the surface (panel (d)). Atmospheric oxygen may be produced via escape during the steam-dominated atmosphere (panel (c)), but this oxygen is drawn down by geological sinks (panel (i)). Volatile cycling is controlled by the rate at which fresh crust is produced (panels (g) and (h)). While melt production continues throughout Venus's history, outgassing is negligible because the mantle is left desiccated after the magma ocean and high pressures mean that volatiles are retained in partial melts (panel (f)). Panel (l) shows water and carbon dioxide in the solid interior as a fraction of total interior and surface reservoirs. The model successfully recovers modern atmospheric ⁴⁰Ar (panel (k)), modern heat flow (panel (e)), surface temperature (panel (a)), and plausible atmospheric ⁴He (panel (k)).

predictions that are consistent with inferred lower atmosphere abundances (Figure 2(k); Chassefiere et al. 1986; Krasnopolsky & Gladstone 2005).

Figure 3 shows all model outputs that successfully reproduce the modern Venusian atmosphere and that possessed surface liquid water for some of Venus's history. Figure 3(a) shows the evolution of surface and mantle temperature, including a transition to temperate surface conditions beginning around ∼10⁷ yr. In these model runs, when Venus has cooled sufficiently such that OLR drops below the runaway greenhouse limit (Figure 3(e)), atmospheric water vapor condenses out of the atmosphere (Figure 3(c)) and onto the surface (Figure 3(d)). Globally averaged ocean depths of a few hundred meters are typical, although there are several model runs with oceans around 1 km deep (Figure 3(d)). While conditions are right for surface liquid water, carbon cycle feedbacks are active (Figure 3(f)). Specifically, the silicate weathering feedback removes carbon dioxide from the atmosphere (Figure 3(c)) and sequesters it in the interior (Figure 3(l)). During this habitable phase, both carbon dioxide and water are returned to the mantle via subduction (plate tectonics) or slab delamination (stagnant lid), which allows for continuous outgassing (Figure 3(f)). On long timescales, outgassing and weathering carbon fluxes are balanced.

Eventually, since a constant albedo is assumed, the secular increase in solar luminosity triggers a transition back into runaway greenhouse. When ASR once again exceeds the runaway greenhouse limit (Figure 3(e)), all surface liquid water is converted to water vapor (Figure 3(d)). Atmospheric carbon dioxide also begins to rise because weathering reactions cease while the outgassing of carbon dioxide may continue. Note that while the return to runaway greenhouse can, in principle, produce another magma ocean if the atmospheric water inventory is large enough, this does not occur for any of the model runs consistent with modern Venus because it is impossible to lose such a large water reservoir subsequently. The duration of surface habitability in Figure 3 ranges from 0.04 to 3.5 Gyr with 95% confidence. More recent habitability is precluded because it is difficult to remove remaining water and oxygen late in Venus's evolution (see discussion).

The evolution of melt production (Figure 3(g)) and crustal depth (Figure 3(j)) are comparable to the nonhabitable scenarios, as are atmospheric ⁴⁰Ar (Figure 3(k)) and ⁴He (Figure 3(l)) evolutions. While oxygen is still being produced from net water loss throughout Venus's history, it is ultimately removed via a combination of crustal and outgassing sinks (Figure 3(i)), leading to an oxygen-free modern atmosphere (Figure 3(c)).

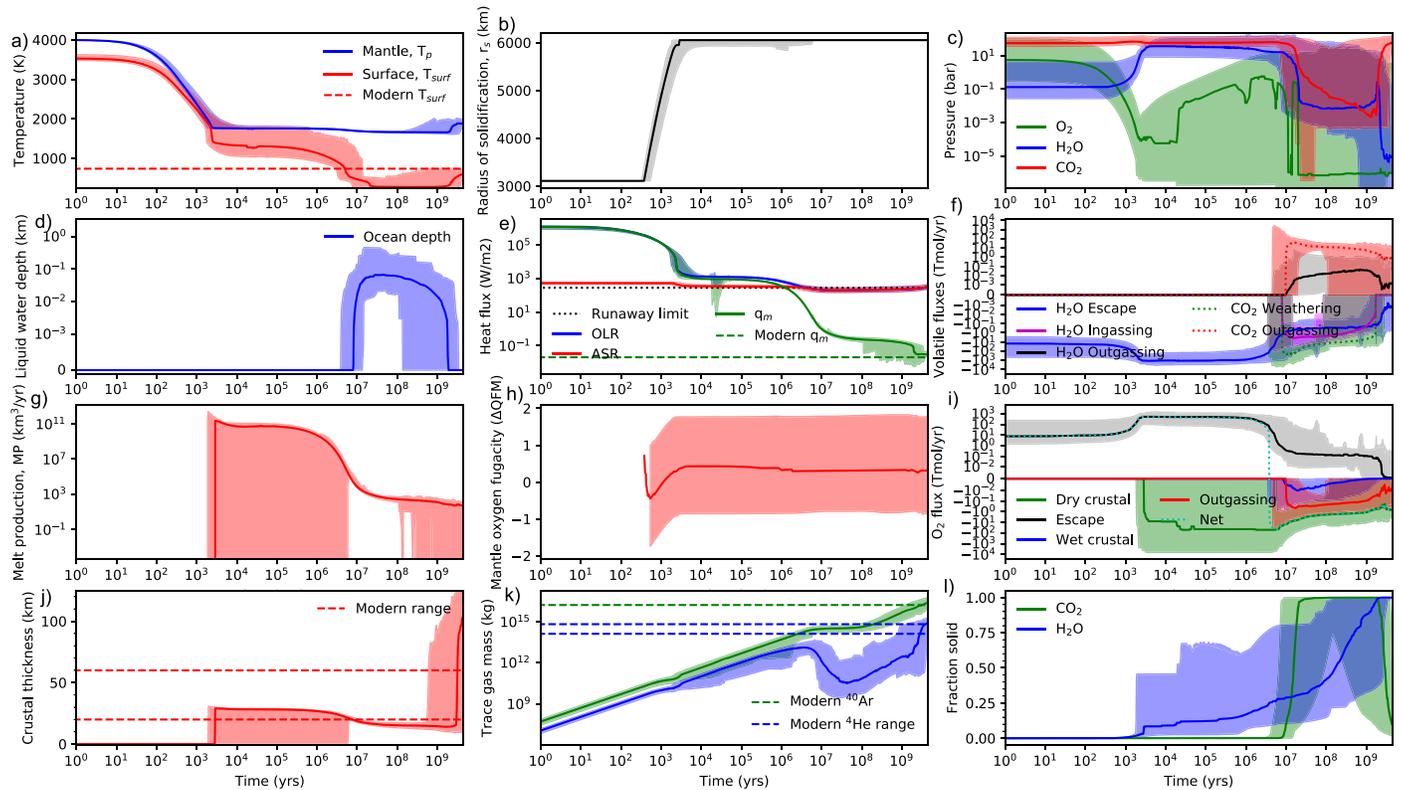

**Figure 3.** Model runs that successfully reproduce modern Venus's bulk atmosphere and where past surface conditions were transiently habitable. The lines are median values, and shaded regions denote 95% confidence intervals. The time evolution spans from post-accretionary magma ocean to the present day ($4.5 \times 10^9$ yr). Panel (a) shows the evolution of mantle potential temperature (blue) and surface temperature (red) over the last 4.5 Gyr. The initial magma ocean (panel (b)) persists from anywhere between a few thousand years to a few million years, depending on initial volatiles, consistent with previous studies (Hamano et al. 2013). The magma ocean ends when Venus's interior cools such that heat flow from the interior drops below the runaway greenhouse limit (panel (e)). When this occurs, atmospheric water vapor (panel (c)) condenses onto the surface, producing a global ocean a few hundred meters deep (panel (d)). A temperate carbon cycle also commences (panel (f)), where volatile cycling is controlled by the rate at which fresh crust is produced (panels (g) and (h)). Oxygen production is low for most of Venus's temperate history because H escape is diffusion limited through a dry upper atmosphere (panel (i)). Eventually, increasing solar luminosity triggers a transition back into runaway greenhouse (panel (e)). When this occurs, surface oceans vaporize (panel (d)), weathering reactions cease (panel (f)), and outgassing returns carbon dioxide to the atmosphere (panel (c)). Panel (l) shows water and carbon dioxide in the solid interior as a fraction of total interior and surface reservoirs. The model successfully recovers modern atmospheric $^{40}$Ar (panel (k)), modern heat flow (panel (e)), surface temperature (panel (a)), and plausible atmospheric $^4$He (panel (k)).

### 3.1. Parameters Controlling Venus's Geochemical Evolution

Next, we consider which variables control whether modern Venusian conditions are recovered, and why only some model runs permit a habitable past. Figure 4 shows circles representing a sample of individual model runs. Magenta circles are the model outputs in Figure 2 that recover the modern Venusian atmosphere and were never habitable ("never-habitable Venus"); yellow circles are the model outputs in Figure 3 that recover the modern Venusian atmosphere and were transiently habitable ("habitable Venus"). Cyan circles show other model runs that do not recover modern atmospheric $CO_2$, $H_2O$, and $O_2$ abundances ("failed Venus"). Bold axes show input values such as initial conditions and other parameter choices, whereas nonbold axes show output values such as final atmospheric abundances after 4.5 Gyr of evolution. Figure 4(a) shows final atmospheric oxygen after 4.5 Gyr as a function of the efficiency of dry crustal oxidation. Note that both never-habitable and transient-habitable Venus scenarios typically require a relatively efficient crustal sink for oxygen, whereby >0.1% of iron in newly produced crust is oxidized. Less efficient crustal sinks leave behind too much atmospheric oxygen.

Figure 4(b) shows the spread of bond albedo values for all model runs. As expected, all successful model runs that were transiently habitable (yellow) have a bond albedo >0.4. These transiently habitable outputs could represent a scenario described in Way & Del Genio (2020), where high albedos due to cloudiness at the substellar point (enabled by Venus's slow rotation) can maintain a temperate surface. There is a negative correlation between albedo and initial water for habitable scenarios. This is because higher albedos imply longer periods of transient habitability, and it is more difficult to remove the leftover water when Venus reenters runaway greenhouse late since the solar XUV flux diminishes exponentially with time. This is also reflected in Figure 4(f), which shows the relationship between albedo and maximum ocean depth. Deep (∼1 km) oceans are only permitted for low bond albedos since this enables earlier and more vigorous water loss while solar XUV fluxes are high.

Figure 4(c) shows what initial water and carbon dioxide inventories are required to reproduce modern Venus conditions, compared to the full range sampled in our model. These results suggest that Venus must have formed with, at most, a few Earth oceans worth of water to be desiccated today (although note that we neglect losses from nonthermal processes such as impact-induced loss). Figure 4(c) also suggests that Venus's initial endowment of carbon was comparable to the modern atmospheric reservoir. Figure 4(d) shows the final water and carbon dioxide evolution of all model runs after 4.5 Gyr of evolution. The broad parameterizations in our model permit a number of evolutionary scenarios that do not recover modern Venus, e.g., planets for which not all water is lost and (high-albedo) planets that remain habitable after 4.5 Gyr with most $CO_2$ sequestered in the mantle (Figure 4(d)). Figures 4(g), (h), and (i) show that results are relatively insensitive to initial mantle redox state, the XUV saturation time of the early Sun, and the radionuclide inventory, respectively. However, it should be noted that very low radionuclide inventories (<0.75 that of Earth) are disfavored for habitable Venus owing to the need for comparatively large and recent oxygen sinks, which are driven by melt production rates. The lack of any XUV evolution dependence is attributable to the most important, late stages of atmospheric loss being diffusion limited rather than XUV limited (see next section). A broader range of initial mantle redox conditions are also considered below.

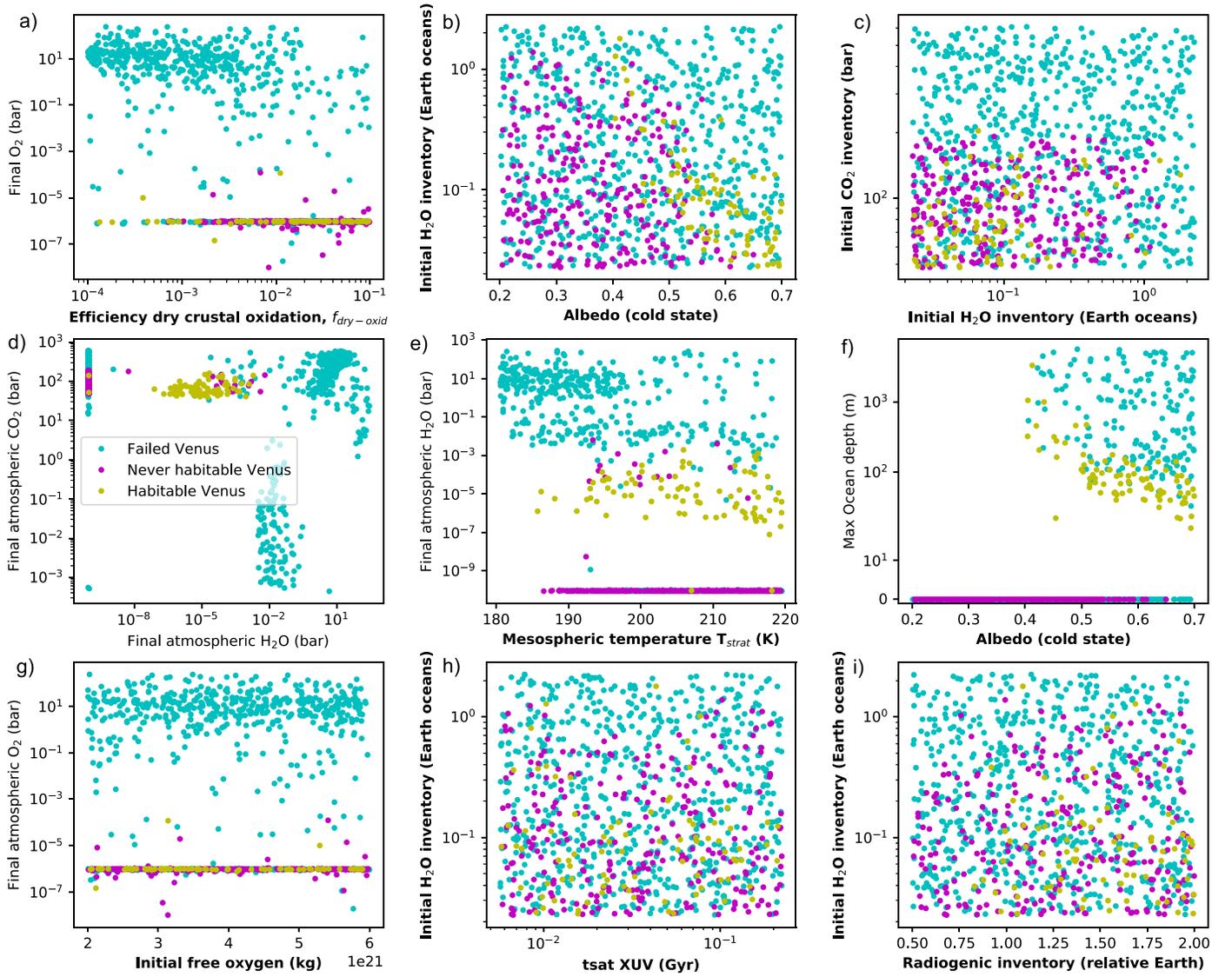

**Figure 4.** Key parameters controlling Venus's atmospheric-interior evolution. Each circle represents a single model run. Magenta circles represent model runs that successfully recovered the modern Venusian atmosphere and were never habitable (Figure 2), yellow circles represent model runs that successfully recovered the modern Venusian atmosphere and were habitable in the past (Figure 3), and cyan circles represent other model runs that did not recover modern atmospheric composition (either too much modern water vapor or molecular oxygen, or too little $CO_2$). Bold axes represent independent variables (initial conditions or input parameters), and nonbold axes represent model outputs. Panel (a) shows the final atmospheric oxygen after 4.5 Gyr as a function of the dry crustal oxidation efficiency; the modern low-oxygen atmosphere implies relatively efficient crustal oxidation over much of Venus's history. Panel (b) shows the assumed bond albedo compared to initial water inventory; transient habitable requires an average albedo exceeding 0.4. Panel (c) shows initial water inventories and initial carbon dioxide inventories, panel (d) shows final atmospheric $CO_2$ inventories and final atmospheric water inventories, panel (e) shows final atmospheric oxygen as a function of upper atmosphere temperature, panel (f) shows peak ocean depth as a function of albedo, panel (g) shows final atmospheric oxygen as a function of initial mantle redox, and panels (h) and (i) show the relationships between initial water inventory, solar XUV saturation time, and initial radiogenic inventory.

### 3.2. Radiative Cooling of the Upper Atmosphere Favors a Habitable Past Climate

Figure 4(e) shows final atmospheric oxygen after 4.5 Gyr as a function of upper atmosphere temperature. Both habitable and nonhabitable past climates are favored when average upper atmosphere temperature exceeds ~190 K. However, the atmosphere of modern Venus between the cloud deck and the homopause typically reaches cooler temperatures, around 150–180 K (Pätzold et al. 2007; Mahieux et al. 2015). This is due, in part, to strong radiative cooling from carbon dioxide (e.g., Wordsworth & Pierrehumbert 2013;

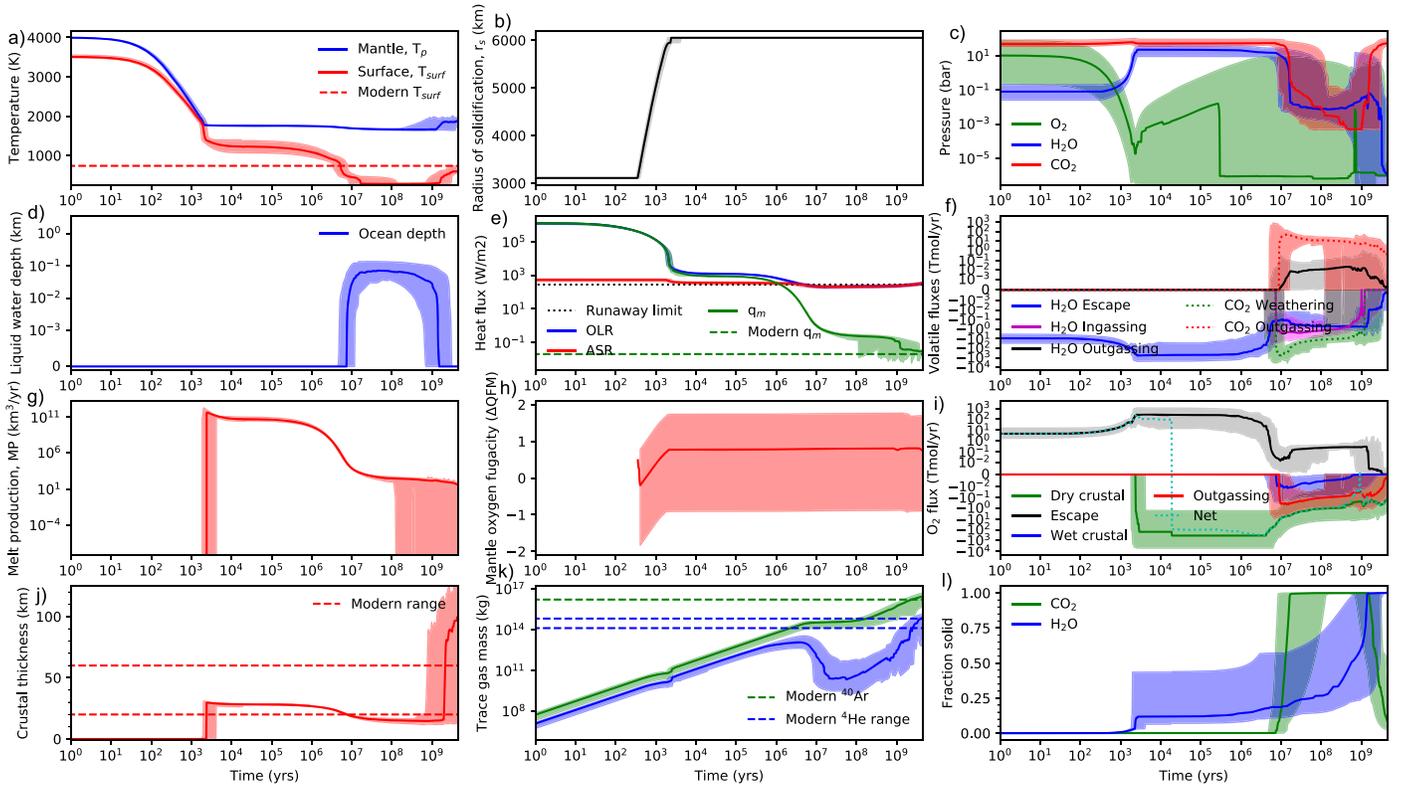

**Figure 5.** Transient habitability is required to recover modern Venus conditions if strong radiative cooling of the upper atmosphere throttles H escape. Nominal calculations were repeated except that we allow the upper atmosphere temperature responds to CO₂ mixing ratio, as described in the text. All model outputs that recover modern Venus conditions are plotted. The lines are median values, and shaded regions denote 95% confidence intervals. Panels are the same as in Figure 3. A period of transient habitability (panels (a) and (d)) allows pCO₂ to be sequestered in the interior via weathering and subduction (panel (f)), thereby enabling greater subsequent water loss and a plausible fit to the modern atmosphere (panel (c)). There are virtually no equivalent never-habitable model outputs that reproduce modern Venus when strong radiative cooling of the upper atmosphere is assumed.

Johnstone et al. 2018). Here, we conduct a sensitivity test where we allow the isothermal upper atmosphere temperature, $T_{meso}$, to vary as a function of carbon dioxide mixing ratio in the upper atmosphere, $f_{CO_2-meso}$:

$$T_{meso} = 214 - 44 \times f_{CO_2-meso}. \quad (1)$$

This is a crude parameterization that ensures Earth-like stratospheric temperatures (214 K) for N₂-O₂ dominated atmospheres and modern Venus-like upper atmosphere temperatures (170 K) for CO₂-dominated upper atmospheres. If anything, this parameterization may underestimate radiative cooling from CO₂; observed minimum temperatures in Venus's homosphere are frequently <170 K (Pätzold et al. 2007; Mahieux et al. 2015), and theoretical calculations of upper atmosphere temperature profiles for varying pCO₂ and stellar insolation also yield cooler temperatures (Wordsworth & Pierrehumbert 2013).

When Equation (1) is adopted and all calculations are repeated, there are virtually no model runs that successfully reproduce modern Venus without transient habitability (the 2/7000 "never-habitable" model runs that do require extremely efficient dry oxidation, ~10%, and even then only recover a 50 bar CO₂ modern atmosphere). This is because carbon dioxide throttles the escape of hydrogen when water concentrations are low. The escape of water is limited by the rate at which hydrogen-bearing species diffuse through the background atmosphere, and cooler temperatures lower water concentrations in the upper atmosphere. Under never-habitable scenarios the atmosphere was continuously CO₂ dominated since accretion owing to the low solubility of carbon dioxide in the magma ocean, and so this CO₂ throttling limits Venus from losing enough water to recover the modern atmosphere.

In contrast, it is possible to recover modern Venus conditions with radiative cooling of the upper atmosphere if transient habitability is permitted. Figure 5 shows all model runs that successfully cover modern Venus conditions when the parameterization described by Equation (1) is adopted. If surface water condenses, then carbon dioxide is weathered out of the atmosphere and sequestered as carbonates (Figures 5(c), (f)). Water loss may then occur during the moist greenhouse phase, when the dominant background gas is nitrogen. Moreover, when the planet reenters the runaway greenhouse, there is a transient steam atmosphere where enhanced water loss may occur. It takes time for CO₂ from outgassing to accumulate to throttle further escape. This dynamic can be seen in Figure 5(c), where atmospheric water vapor increases from the evaporating ocean before pCO₂ is massively degassed, thus providing a short window for water loss. This sequence of events can also be seen in Figure 6, which depicts the "habitable Venus" outputs from Figure 5, as well as a selection of failed Venus model outputs that never recover modern conditions owing to water loss being throttled by strong radiative cooling.

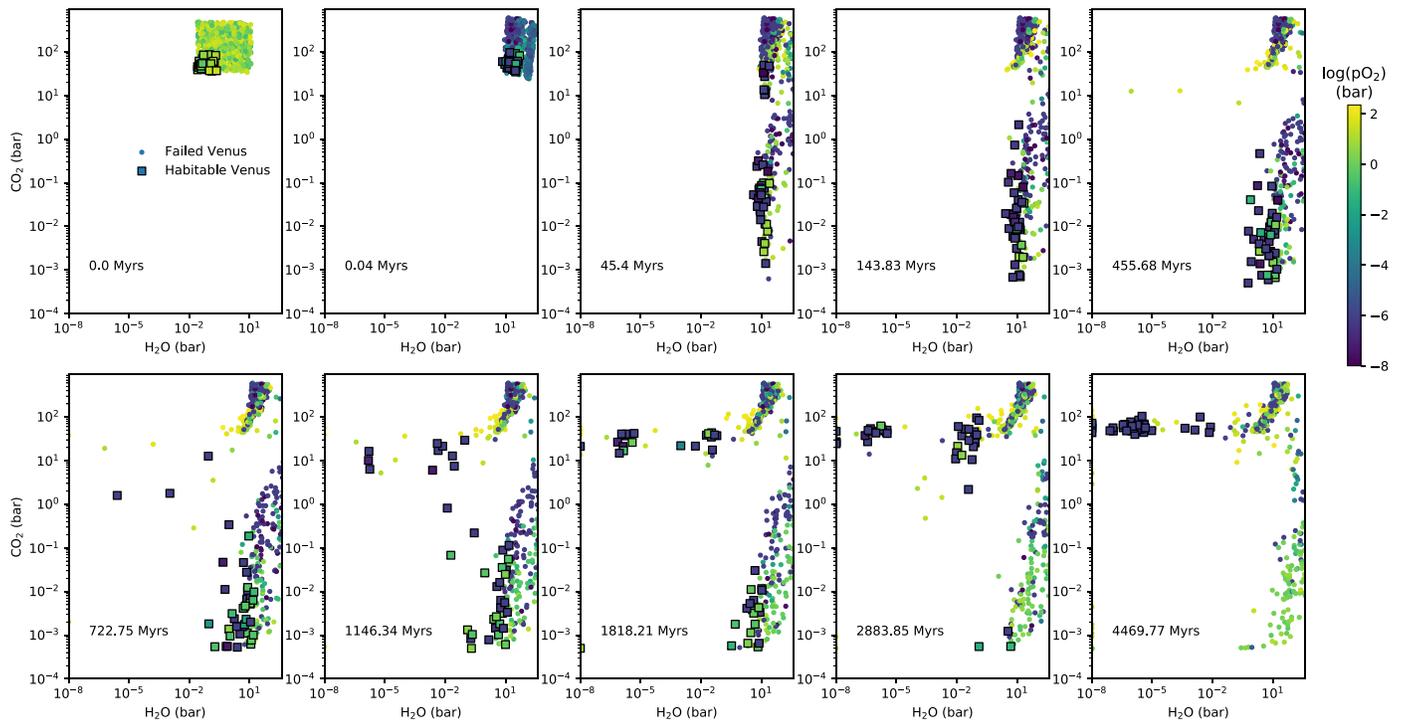

**Figure 6.** Transient habitability is needed to recover modern Venus if strong radiative cooling of the upper atmosphere throttles H escape. The time evolutions (in Myr) of atmospheric carbon dioxide, water, and oxygen are shown for the outputs in Figure 5, denoted here by squares. A number of other model runs that fail to recover modern Venus conditions are shown as circles. The x-axes and y-axes denote atmospheric water vapor and carbon dioxide, respectively, whereas the color bar shows atmospheric oxygen. Note that there are no "never-habitable" model runs that recover modern Venus in this figure because radiative cooling throttles escape. Only by passing through a transient-habitable state whereby $pCO_2$ is lowered by silicate weathering can upper atmosphere temperatures increase such that sufficient water can escape to explain the modern desiccated atmosphere. The comparatively few circles for which water is lost have too much leftover oxygen to match the observed atmosphere. An animation of the time evolution is available in the online Journal. The animation spans the full time shown in the static figure (0–4469.77 Myr) and progressing at a variable time step.

(An animation of this figure is available.)

## 4. Discussion

The most important result to emerge from these calculations is that both a habitable Venusian past and a nonhabitable Venusian past can be broadly reconciled with modern constraints. Either scenario may produce a $CO_2$-dominated atmosphere with negligible oxygen and water vapor, modern interior heat flow, and atmospheric $^{40}Ar$ and $^4He$ abundances that are consistent with observations. This emphasizes the need for in situ observations to better constrain past atmospheric evolution and surface conditions. The upcoming possibility of characterizing the atmospheres of Venus-like exoplanets provides additional motivation for understanding Venus's evolution (Ehrenreich et al. 2012; Kane et al. 2019). Simulated observations suggest that the James Webb Space Telescope may be capable of constraining the atmospheric composition (Barstow et al. 2016; Lustig-Yaeger et al. 2019a), and perhaps even the atmospheric isotopologues (Lincowski et al. 2019), of exo-Venus analogs such as Trappist-1b, c, and d, GJ 1132b, and LHS 3844b. Ultimately, a larger sample of TESS exo-Venus candidates may be characterizable with Webb (Ostberg & Kane 2019), although high-altitude aerosols could preclude definitive assessment of total atmospheric inventories (Lustig-Yaeger et al. 2019b). Figure 4(d) shows that, in addition to the never-habitable and transiently habitable histories that are the focus of this study, dramatically different modern states are permitted by our model. For example, continuously habitable planets or water-rich runaway greenhouses with widely variable atmospheric oxygen levels are all self-consistent model outcomes after 4.5 Gyr of evolution (Figures 4(a), (d)). Understanding why Venus avoided these outcomes and constraining the climate evolution trajectory it underwent to reach its observed state will help inform the interpretation of exo-Venus analog atmospheres. Conversely, the population level data on exo-Venus analogs expected from upcoming observations will help inform reconstructions of Venus's atmospheric evolution.

### 4.1. Sensitivity of Results to Ion Escape, Mantle Depletion, Magma Ocean Volatile Partitioning, and Initial Mantle Redox

Our results are largely insensitive to model assumptions about escape and the evolution of volatile reservoirs during the magma ocean. In the nominal model outputs shown in Figures 2 and 3, we consider only XUV-driven hydrodynamic escape of H (with O and $CO_2$ drag) and diffusion-limited H escape. The diffusion limit to H escape applies to both thermal and nonthermal escape processes. However, it is likely that nonthermal oxygen ion escape also contributed significantly to Venus's atmospheric evolution (Kulikov et al. 2006; Gillmann et al. 2020; Persson et al. 2020). To test the effect of this, we conducted a sensitivity test where we imposed O ion escape estimates from Kulikov et al. (2006; their case 2b). The results of these calculations are shown in full in the Appendices (Figure B1), but in summary, we find that O ion escape does not change our qualitative conclusions. This is because ion escape becomes small after 3.5 Ga, and so it does not help remove excess oxygen from late water loss; efficient crustal sinks and comparatively small initial water reservoirs are still required.

In our nominal calculations, the silicate solidus depends on pressure overburden but is independent of composition. Under a stagnant lid regime, melt accumulation in the lid may progressively deplete the mantle, resulting in an increasing solidus. Melt production—and therefore oxygen sinks—may thus decline over time. To test the impact of mantle depletion, we repeated our calculations allowing for the evolution of the solidus with melt extraction from the mantle. This is described in full in Appendix B.1, and results are shown in Figures B2 and B3. Although the modern average melt production in these sensitivity tests is around half that of the nominal model (Figures B2(g), B3(g)), the habitable and never-habitable model outputs are otherwise unchanged, and the proportions of model runs that successfully recover modern Venus are also indistinguishable from nominal calculations. This is because while decreasing melt production lessens the size of oxygen sinks, it also limits the resupply of water via degassing, which helps recover the desiccated modern Venus atmosphere.

The magma ocean phase in the nominal model terminates with virtually all carbon (and most water) residing in the atmosphere owing to the low solubility of carbon dioxide in silicate melts (Figures 2(l), 3(l)). This is consistent with comparable magma ocean evolution models which assume that volatile partitioning is governed by melt solubility (Lebrun et al. 2013; Schaefer et al. 2016; Salvador et al. 2017). However, volatiles could become trapped within the mantle during magma ocean solidification owing to compaction within the moving freezing front (e.g., Hier-Majumder & Hirschmann 2017). To test the impact of such volatile retention, we repeated our nominal calculations accounting for melt trapping during magma ocean solidification (see Appendix B.2). We adopt a compaction timescale of ~1 Myr, which results in much higher retained volatile fractions than assuming constant melt fraction trapping or merely accounting for equilibrium partitioning of volatiles into cumulates (Elkins-Tanton 2008; Hier-Majumder & Hirschmann 2017; Nikolaou et al. 2019); this sensitivity test represents an end-member case with maximum volatile retention during magma ocean solidification. Figure B4 shows all the never-habitable model outputs from this sensitivity test, and Figure B5 shows all the transiently habitable model outputs from this sensitivity test. In these cases, approximately 70% of total water and 6% of total $CO_2$ resides in the solid mantle after magma ocean solidification (Figures B4(l), B5(l)). Outgassing fluxes are enhanced in both cases (Figures B4(f), B5(f)), but outputs are otherwise comparable to Figures 2 and 3, respectively. This demonstrates that a desiccated initial mantle is not required to recover modern Venus. One key difference, however, is that habitable scenarios are far less frequent: there are 85% fewer habitable model runs compared to nominal calculations. This is because when ~70% of the initial water inventory is retained in the mantle, comparatively large water outgassing fluxes continue for several Gyr, and so it is difficult to explain the low modern atmospheric water abundance. In contrast, never-habitable scenarios are unaffected because (i) the pressure overburden of a ~100 bar $CO_2$ atmosphere inhibits $H_2O$ exsolution from partial melts, resulting in less atmospheric $H_2O$ replenishment, and (ii) never-habitable Venuses are typically desiccated earlier in their history compared to transiently habitable Venuses, which undergo a more recent runaway greenhouse.

The range of initial mantle redox states in nominal calculations was chosen to recover an Earth-like QFM modern mantle. While there is some evidence for a strongly oxidized Venus surface (Fegley et al. 1997), Venus's mantle redox has never been measured. More reducing initial redox states are possible (Armstrong et al. 2019; Sossi et al. 2020), and so all calculations were repeated with a lower free oxygen initial reservoir: (0.3–1.5) × $10^{21}$ kg free O compared to (2.0–6.0) × $10^{21}$ kg in the nominal model. This ensures a modern mantle redox state closer to the iron–wüstite (IW) buffer. Figure B6 shows the model outputs from this sensitivity test that recover modern Venus and were never habitable, whereas Figure B7 shows the outputs that recover modern Venus and experienced transient habitability. Decreasing mantle redox has various impacts on geochemical evolution. More reduced partial melts tend to produce more reducing gases, but carbon also preferentially partitions into solid graphite under reducing conditions lowering total outgassing fluxes. These two effects partially offset one another, and the overall impact on oxygen sinks is small. Additionally, lowered carbon outgassing due to graphite saturated conditions makes it challenging to restore a ~90 bar $CO_2$-dominated atmosphere via outgassing after $CO_2$ drawdown during a habitable epoch, and so there are slightly fewer "habitable" model runs than in nominal calculations (27% fewer). Otherwise, lowering initial mantle redox has a minor effect on Venus's geochemical evolution, with the additional caveat that our model does not permit CO-dominated atmospheres and instead assumes that all outgassed CO is photochemically oxidized to $CO_2$ (see discussion below).

### 4.2. Sensitivity to Atmospheric $N_2$ Abundance

Modern Venus's atmosphere contains ~3 bar $N_2$, but there is considerable uncertainty in the evolution of this inventory, especially if Venus experienced temperate surface conditions. The evolution of atmospheric nitrogen ought to have negligible impact on the "never-habitable" scenarios: when the atmosphere is always dominated by carbon dioxide and water vapor, the comparatively small amount of background $N_2$ will not impact atmospheric escape substantially. However, if Venus did experience transient habitability, then the background nitrogen inventory could have impacted planetary redox evolution. This is because, in the absence of a thick $CO_2$ atmosphere, $N_2$ is the primary noncondensable species that forms the cold trap, inhibiting water loss and oxygen accumulation (Wordsworth & Pierrehumbert 2014; Kleinböhl et al. 2018). Predictive modeling of nitrogen atmospheric evolution on temperate planets is challenging, however. The evolution of Earth's atmospheric nitrogen inventory is uncertain owing to lack of constraints on abiotic fixation pathways and the post-accretion partitioning of nitrogen between the silicate mantle and surface volatile reservoirs. (Stüeken et al. 2016; Wordsworth 2016; Johnson & Goldblatt 2018; Lammer et al. 2019). In our nominal calculations we assumed an Earth-like inventory of 1 bar $N_2$ atmospheric nitrogen that is constant throughout Venus's evolution, but note that efficient abiotic nitrogen fixation could yield enhanced H loss and oxygen accumulation. On the other hand, efficient nitrogen drawdown would enable more rapid water loss during runaway greenhouse, potentially allowing deeper global oceans to be compatible with modern conditions.

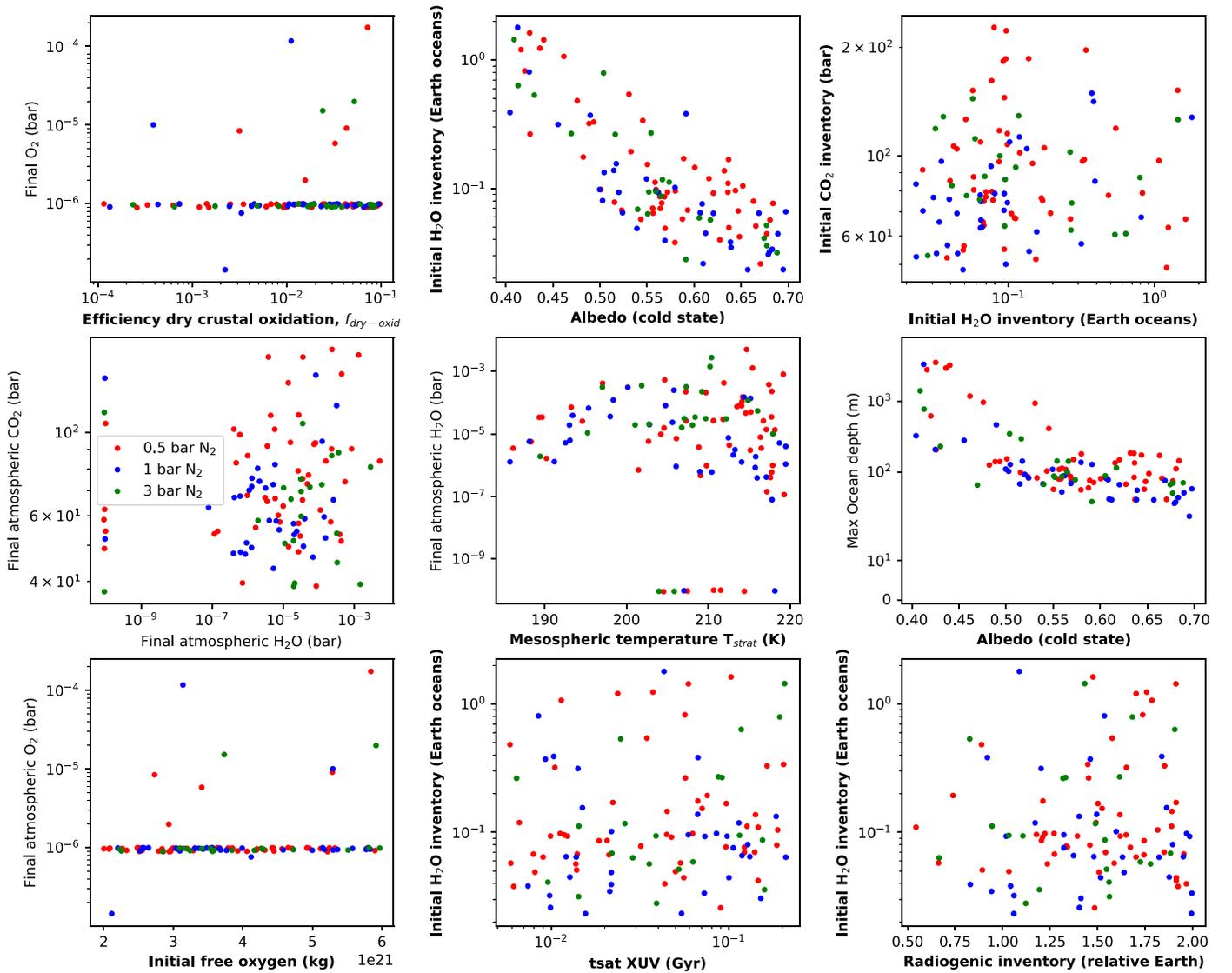

**Figure 7.** Habitable Venus scenarios are largely unaffected by background $N_2$ pressure. Each circle represents a model run that successfully recovers modern Venus conditions and experienced transient habitability. Red circles assume a constant background $N_2$ pressure of 0.5 bar, blue circles 1 bar (nominal model), and green circles 3 bar. Panels and axes are the same as in Figure 4: bold axes represent independent variables (initial conditions or input parameters), and nonbold axes represent model outputs. Lower background $N_2$ pressure results in slightly more model runs that recover modern Venus with transient habitability (H escape more rapid), but the trends and relationships between unknown parameters are unchanged.

To test the impact of possible $N_2$ changes on our results, we performed sensitivity tests with background $N_2$ abundances of 3 and 0.5 bars. Figure 7 shows results from these sensitivity tests where each circle is an individual model run that recovered modern Venus conditions and experienced past habitability. Red circles represent calculations where a 0.5 bar constant $N_2$ background was assumed, blue circles 1 bar $N_2$, and green circles 3 bar. Bold axes show input values such as initial conditions and other parameter choices, whereas nonbold axes show output values such as final atmospheric abundances after 4.5 Gyr of evolution (axes identical to Figure 4). While the parameter values required for a habitable Venus do not change substantially with atmospheric pressure (Figure 7), increasing atmospheric pressure to 3 bars decreases the number of habitable Venus model runs by around 40%, whereas decreasing atmospheric pressure to 0.5 bars increases the number of habitable Venus model runs by 56%. As expected, there is no change in the number of model runs that successfully recover modern Venus without experiencing transient habitability when $N_2$ partial pressure is varied (not shown).

### 4.3. Evidence for a Habitable Ancient Venus?

Our model provides tentative evidence for a habitable Venusian past owing to the $CO_2$ throttling of hydrogen escape in nonhabitable scenarios. If radiative cooling of the upper atmosphere is incorporated into our escape parameterization, then never-habitable model runs cannot successfully reproduce modern Venus's lack of atmospheric water vapor and oxygen (Figures 5, 6). Additionally, escape fluxes are potentially overestimated since we assume that water molecules are completely dissociated to atomic hydrogen and oxygen prior to diffusive transport (e.g., Catling & Kasting 2017). Real escape fluxes could be even lower if hydrogen is retained in water molecules during diffusive transport through a $CO_2$–$N_2$ background atmosphere (e.g.,

Wordsworth & Pierrehumbert 2013). This would make nonhabitable scenarios even less likely to recover modern Venus conditions owing to the additional throttling of water loss (see Appendix A.6 for escape assumptions).

There are several important caveats to this result, however. Crucially, for never-habitable model runs, water is the only hydrogen-bearing species in our model that can escape. Other hydrogen-bearing species, such as $H_2$ and $CH_4$, are not limited by a cold trap and so could potentially provide a vehicle for higher H-escape fluxes. For example, high concentrations of biogenic methane may have allowed for elevated H-escape fluxes on the early Earth (Catling et al. 2001). While we do account for H escape from $H_2$ produced by serpentinization reactions ($2FeO + H_2O \rightarrow 2FeO_{1.5} + H_2$), these reactions are only assumed to occur when liquid water is in contact with newly produced crust (see Appendix A.5). Whether it is possible to continuously generate massive amounts of molecular hydrogen via reactions between supercritical water and rock is unknown. Reactions between supercritical water and silicates are typically assumed to be kinetically limited by sluggish solid-state diffusion and therefore negligible (Zolotov et al. 1997). While there is some experimental evidence for rapid oxidation of volcanic glass under wet, hot conditions (Berger et al. 2019), it is difficult to extrapolate these results to planetary scales.

Another possible mechanism for $H_2$ and $CH_4$ generation is via large impacts, which may transiently convert surface water inventories to reduced gases via reactions with metallic and ferrous iron impactor material (Zahnle et al. 2020). Future work could investigate whether impact-induced water reduction is sufficiently efficient, and if the resulting $CH_4/H_2$ is sufficiently long-lived, to modify our conclusions on Venus's escape history.

A key limitation to the model used in this study is that it is restricted to $CO_2$–$H_2O$-dominated atmospheres. The primordial magma ocean atmosphere of Venus may have been predominantly CO and $H_2$ rather than $CO_2$ and $H_2O$ dominated (Sossi et al. 2020). If this CO–$H_2$ atmosphere was sufficiently long-lived, then Venus could have lost much of its hydrogen inventory rapidly because molecular $H_2$ is noncondensible. One promising avenue for future work would be to model the coevolution of $CO_2$–CO–$H_2$–$H_2O$–$CH_4$ atmospheres and the magma ocean, including self-consistently calculating the thermal structure of the upper atmosphere, to better constrain H-escape fluxes during early evolution. While a more reducing initial atmosphere would enhance early escape fluxes, without such explicit modeling, it is unclear whether this scenario would remove the need for transient habitability. This is because, as H abundances declined, the resulting CO-dominated atmosphere would provide a weak greenhouse effect compared to $CO_2$- and $H_2O$-dominated atmospheres. Once surface temperatures cooled below ~900 K, the remaining hydrogen would react with CO to form water and $CO_2$ (Sossi et al. 2020), which would be subject to the same cold-trap throttling as described above. Our calculations sample initial water inventories as low as 2% of an Earth ocean, which is a reasonable representation of a scenario whereby most hydrogen is lost in a CO–$H_2$ magma ocean atmosphere. And yet even this comparatively modest residual water reservoir cannot be completely lost to space without a transient-habitable interval.

While transient habitability is favored when $CO_2$-rich atmospheres are assumed to throttle escape by cooling the upper atmosphere (Section 3.2), greater retention of volatiles in the mantle during magma ocean solidification lowers the number of transiently habitable model outputs (Section 4.1). In particular, if ~70% (Hier-Majumder & Hirschmann 2017) of the initial water inventory is retained in the solid mantle, then "habitable Venus" scenarios are more difficult to attain (85% less frequent) owing to late outgassing replenishment of $H_2O$ that cannot be easily lost to space. It should be noted, however, that our outgassing model is highly simplified. It ignores H-bearing sulfur species, and it does not allow for episodic melt production or variable extrusive/intrusive melt emplacement. A more realistic outgassing model could potentially reconcile habitability and mantle volatile retention. When both $CO_2$ throttling of escape and substantial volatile retention are assumed, virtually no model runs recover modern Venus, whether transiently habitable or never habitable.

Both nonhabitable and habitable scenarios tend to favor high recent melt production, with roughly tens of $km^3 \, yr^{-1}$ being typical values after 4.5 Gyr of evolution. Mean melt production in the last 0.5 Gyr for habitable model outputs is 7–97 $km^3 \, yr^{-1}$ (95% confidence) and 0–95 $km^3 \, yr^{-1}$ (95% confidence) for never-habitable model runs. Figure 8 shows the evolution of melt production for a subset of never-habitable (top row) and habitable (bottom row) nominal model runs that successfully recover modern Venus. To illustrate how tectonic regime evolution controls melt production, these have been decomposed into model runs where plate tectonics ended prior to 3.5 Ga (right column) and model runs where plate tectonics ended after 1.5 Ga (left column). The transition from plate tectonics to stagnant lid causes a transient decrease—and sometimes a temporary cessation—in melt production as mantle temperature increases. While recent melt production is typically high, negligible melt production in the last 0.5 Gyr is still permissible, especially if plate tectonics ended recently (Figures 8(a), (c)). Broadly speaking, high crustal production rates are needed to remove leftover oxygen from recent water loss, although modern melt production may be negligible in model runs where oxygen was removed completely prior to ~0.5 Ga. High melt production may not be reflected in modern resurfacing rates, which are likely <2 $km^3 \, yr^{-1}$ (Head et al. 1991). Specifically, melt production could be largely intrusive and/or episodic. Our habitable model outputs illustrate the self-consistency of a scenario proposed by Way & Del Genio (2020): recent habitability could have ended with enhanced magmatic production that both caused recent resurfacing and provided an oxygen sink for substantial amounts of $O_2$.

In summary, while a habitable past is tentatively favored by our model, there are several important caveats on this conclusion that require further study. Additionally, this model is agnostic on the dynamical plausibility of transitioning from a post-accretion runaway greenhouse to the habitable, high-albedo states modeled in GCMs (Way et al. 2016; Way & Del Genio 2020). It should also be noted that of the ~10% of nominal model runs that successfully reproduce modern Venus, around 9% are never habitable and only ~1% experience transient habitability. While this does not necessarily reflect the relative probability of such scenarios because parameter ranges for initial inventories are arbitrary and albedo feedbacks are omitted, it does show that a degree of parameter tuning is required for habitable Venus histories.

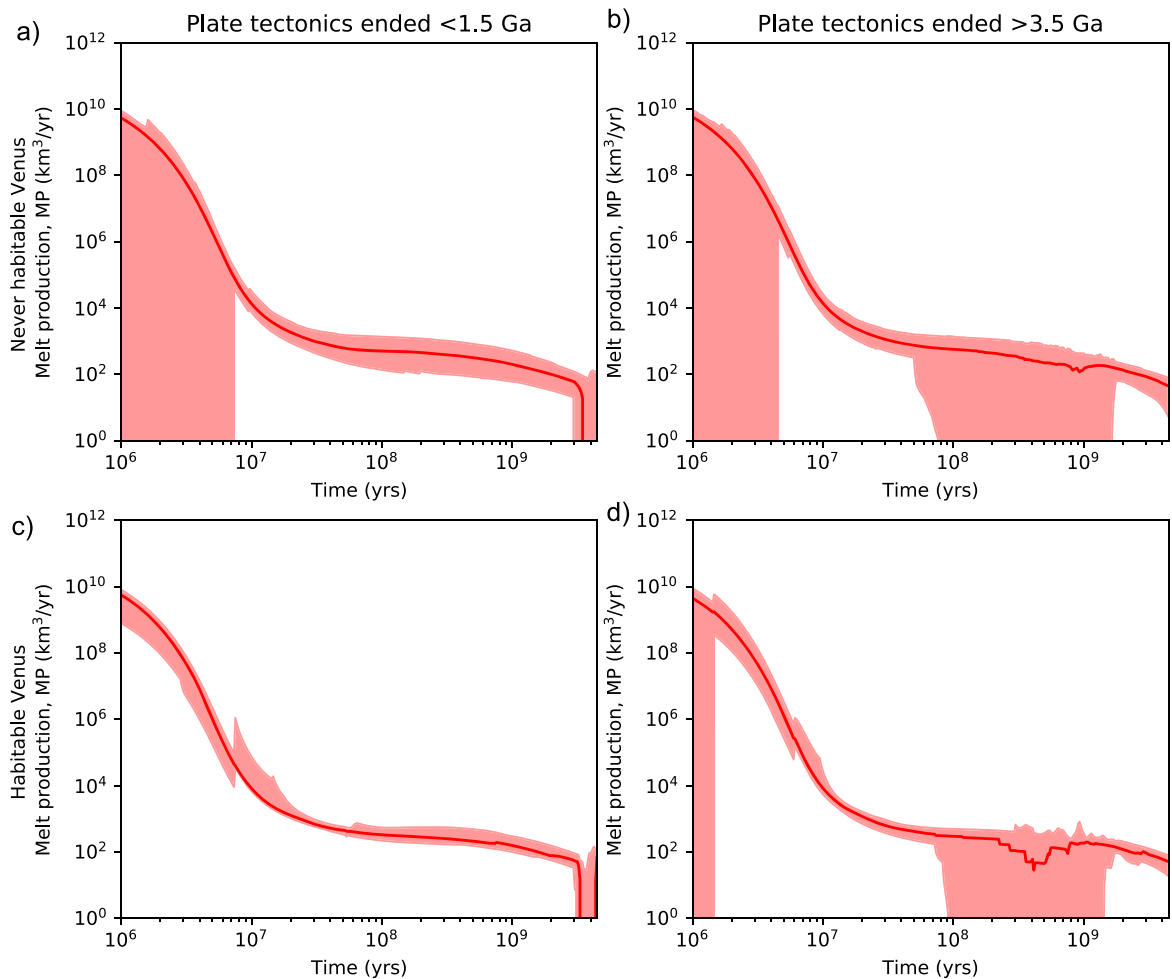

**Figure 8.** Melt production evolution for both habitable (Figure 3) and never-habitable (Figure 2) model scenarios decomposed by how recently plate tectonics ended. The red lines and shaded regions denote median outputs and 95% confidence intervals, respectively. Panels (a) and (b) show never-habitable model outputs that recover modern Venus, whereas panels (c) and (d) show habitable outputs that recover modern Venus. The left column denotes all model runs where the transition from plate tectonics to stagnant lid occurred late, <1.5 Ga, and the right column denotes all model runs where the transition occurred early, >3.5 Ga. The transition from plate tectonics to stagnant lid causes a temporary decrease in mantle production until the mantle heats up such that melting below the lid may resume. Melt production for model runs when plate tectonics ended between 3.5 and 1.5 Ga are not shown but are intermediate to these two extremes.

### 4.4. Duration of Habitable Surface Conditions and Tectonic Regime Transitions

If Venus was habitable in the past, our model constrains the duration of that habitability to around 0.04–3.5 Gyr (95% confidence, median 1.9 Gyr), implying that Venus's surface has been uninhabitable since at least 1.0 Ga. While broadly consistent with purely climatological calculations that permit surface habitability persisting until resurfacing around 0.7 Ga (Way et al. 2016), this is not a tight constraint. The permitted surface water reservoir does inversely correlate with the duration of habitability, however. Figure 9(a) shows the maximum global ocean depth as a function of the duration of surface habitability for model runs that successfully reproduce modern Venus, as well as a sample of "failed" model runs that do not recover modern pCO₂, pO₂, and pH₂O. Deep global oceans (∼km) are only permitted if habitability is lost within the first Gyr, when solar XUV fluxes were comparatively high. The only conditions where habitability persisted until more recently are for global ocean depths of ∼100 m. This is because recent habitability with large surface reservoirs makes it difficult to subsequently remove water and oxygen to recover the modern atmosphere. Removing water is typically more difficult than removing oxygen. Indeed, many "failed Venus" model outputs resume a magma ocean phase when they reenter runaway greenhouse, thereby removing all molecular oxygen via reactions with the molten surface. But these planets are unable to lose the large amounts of water required to induce a second magma ocean.

Model outputs also provide crude constraints on the tectonic history of Venus. Figure 9(b) shows the modern ⁴⁰Ar(atmosphere)/⁴⁰Ar(total) ratio as a function of the transition time between plate tectonics and a stagnant lid regime. The actual ratio, using observed ⁴⁰Ar(atmosphere) = $1.6 \times 10^{16}$ kg, is overplotted for comparison. The best fit to observed ⁴⁰Ar is achieved if Venus experienced at least 0.5 Gyr of plate tectonics before transitioning to a stagnant lid regime. This is because an early stagnant ensures high early mantle temperatures and larger integrated melt production over Venus's history. However, our mantle convection model uses simple parameterizations for plate tectonics and stagnant lid regimes that may not necessarily capture episodic bursts of melt production. Coupling more comprehensive mantle convection models (e.g., Noack et al. 2012;

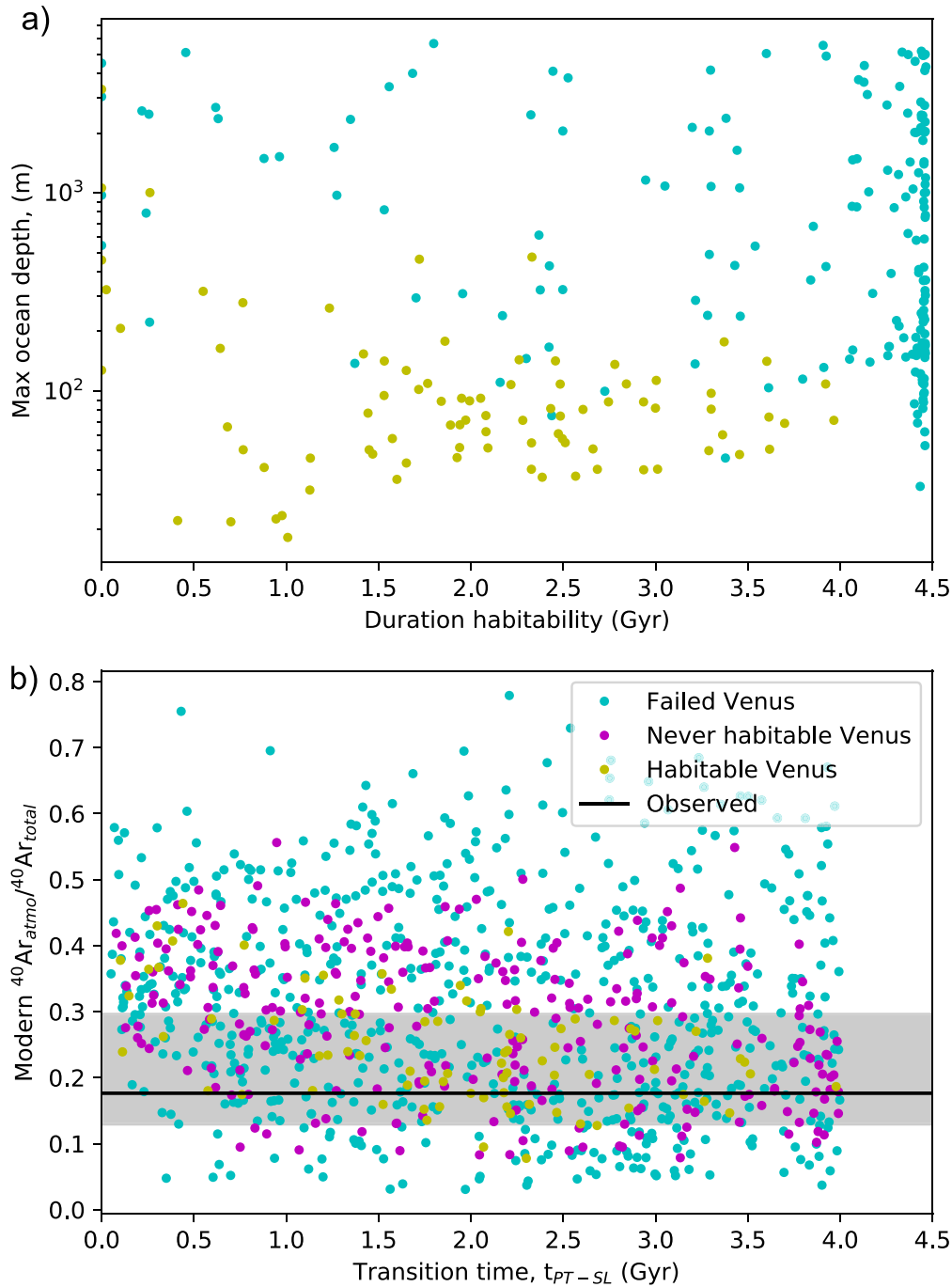

**Figure 9.** Duration of habitability and tectonic regime constraints. Each circle represents a single model run. Magenta circles represent model runs that successfully recovered the modern Venusian atmosphere and were never habitable (Figure 2), yellow circles represent model runs that successfully recovered the modern Venusian atmosphere and were habitable in the past (Figure 3), and cyan circles represent other model runs that did not recover modern atmospheric composition. Panel (a) shows peak global ocean depth as a function of the duration of habitable surface conditions. Recent habitability (in the last ∼2 Gyr) is only compatible with shallow global oceans owing to limits to subsequent water loss. Panel (b) shows the atmospheric $^{40}$Ar relative to total $^{40}$Ar after 4.5 Gyr compared to the transition time between plate tectonics and a stagnant lid; a late transition is slightly easier to reconcile with the observed atmosphere (black and gray shaded).

Gillmann et al. 2020) to planetary climate–redox evolution is one possible avenue for future work.

### 4.5. Implications for Future Venus Missions

Future in situ observations could resolve whether Venus was ever habitable. Atmospheric measurements of Xe and Kr isotopic inventories by the upcoming DAVINCI mission (Garvin et al. 2020) will better constrain the escape history of Venus and, by extension, inform understanding of past climate evolution (Baines et al. 2013; Kane et al. 2019; Zahnle et al. 2019). DAVINCI is also expected to provide more precise constraints on bulk atmospheric D/H (Garvin et al. 2020). One promising avenue for future modeling efforts is to incorporate realistic models of D/H fractionation (under different escape regimes and atmospheric compositions) into whole-planet geochemical evolution models such as PACMAN that include deep volatile cycling and exogeneous supply. Such modeling efforts will be necessary to interpret improved D/H constraints.

Among the nominal model outputs that experienced a habitable past (Figure 3), globally averaged ocean depth peaked at around 100 m. While it is difficult to predict ocean cation and anion abundances owing to unknown contributions from weathering, hydrothermal, and clay-forming processes, if an Earth-like ocean composition is assumed for illustrative purposes, then our model predicts evaporite deposits of around $\sim 10^{18}$ kg halite ($\sim 1$ m global equivalent layer thickness) and $\sim 10^{17}$ kg anhydrite ($\sim 0.1$ m global equivalent layer thickness). Peak surface temperature during and after ocean evaporation is typically <800 K (81% of habitable model outputs), which is below the melting points of both halite and anhydrite (Schofield et al. 2014). Calcite deposits are not stable with respect to sulfatization under modern surface conditions, and they are also more likely to volatilize during metamorphic heating (Zolotov 2019). However, the possibility of calcite deposits beneath impermeable strata cannot be excluded. While assuming that all dissolved cations and anions ultimately form evaporites during ocean evaporation is an oversimplification, the size of the predicted deposits is potentially comparable to that of Earth's evaporites (Warren 2010) and so could be accessible to remote sensing via NIR imaging by the upcoming VERITAS (Smrekar et al. 2020) and EnVision (Widemann et al. 2020) missions. Surface regions conjectured to have survived resurfacing, such as tessera terrain (Hansen & López 2010), are promising locations to search for evaporite deposits and other evidence of hydration.

## 5. Conclusions

1. A coupled mode of Venus's atmosphere–interior–climate evolution implies that both habitable and never-habitable histories are consistent with modern constraints. Both scenarios can reproduce a dense $CO_2$-dominated modern atmosphere with negligible water vapor and reasonable abundances of outgassed $^{40}$Ar and $^4$He. Which of these histories was realized depends on cloud albedo feedbacks.

2. Habitable past scenarios are favored if $CO_2$-rich atmospheres radiatively cool the mesosphere. This is because water escape is throttled by $CO_2$ in nonhabitable scenarios, and so it is difficult to obtain a modern atmosphere with ∼90 bar $CO_2$ and negligible water vapor. Habitable histories avoid this problem by temporarily sequestering $CO_2$ in carbonates, thereby permitting elevated $H_2O$ escape when their atmospheres reenter moist-to-runaway greenhouse phases before degassed $CO_2$ is returned to the atmosphere.

3. On the other hand, habitable Venus outcomes are less frequent when volatiles are assumed to be mostly retained in the solid mantle during magma ocean solidification. This is because water recently outgassed cannot be easily lost to space rapidly enough to explain the observed low $H_2O$ abundance.

4. In our habitable past scenarios, Venus's surface was temperate for 0.04–3.5 Gyr (95% confidence), during which time the global ocean depth may have been a few hundred meters. Assuming plausible ocean cation abundances, a surface ocean of this size ought to have left behind sizable evaporite deposits accessible to future observations.

We thank Giada Arney, Rory Barnes, Rudy Garcia, Michael Way, and Nicholas Wogan for helpful conversations and suggestions. We also thank Stephen Kane and an anonymous reviewer for their constructive comments. J.K.-T. was supported by NASA through the NASA Hubble Fellowship grant HF2-51437 awarded by the Space Telescope Science Institute, which is operated by the Association of Universities for Research in Astronomy, Inc., for NASA, under contract NAS5-26555.

## Appendix A
## Supplementary Methods

The PACMAN model used in this study is a modified version of that described in Krissansen-Totton et al. (2021). This appendix describes the new features added to this model. The Python code for both versions of the model is publicly available on GitHub.[5] Table A1 shows the parameter ranges assumed in nominal model calculations.

### A.1. Thermal Evolution

Planetary thermal evolution is specified by energy budget and temperature-dependent viscosity. The time evolution of mantle potential temperature, $T_p$ (K), is determined by the following equations. During magma ocean, thermal evolution is governed by the following equation:

$$\frac{4\pi \rho_m c_p (r_p{}^3 - r_s{}^3)}{3} \frac{dT_p}{dt} = \frac{4\pi \rho_m q_r (r_p{}^3 - r_c{}^3)}{3}$$
$$- 4\pi q_m r_p{}^2 + Q_c + 4\pi \rho_m H_{\text{fusion}} r_s{}^2 \frac{dr_s}{dt}. \quad (2)$$

Here $q_r$ (W kg$^{-1}$) is the heat production via radiogenic isotopes. Venus's initial endowment of $^{235}$U, $^{238}$U, $^{232}$Th, and $^{26}$Al is sampled from a range of 0.5–2.0 times Earth's initial radiogenic inventory. Venus's modern K/U silicate ratio is uniformly sampled from 6000 to 8440 (Kaula 1999). Additionally, $\rho_m = 4000$ kg m$^{-3}$ is the assumed average density of mantle material, $r_p$ is the total radius of the planet, $r_c$ is the radius of the core, $r_s$ is the radius of solidification, $H_{\text{fusion}} = 4 \times 10^5$ J kg$^{-1}$ is the latent heat of fusion of silicates, and $c_p = 1200$ J kg$^{-1}$ K$^{-1}$ is the heat capacity of silicates. The time evolution of the radius of solidification is calculated using the same approach as that described in Krissansen-Totton et al. (2021). Heat flux from the metallic core, $Q_c$, is parameterized as an exponential decay toward a modern flux consistent with the absence of a dynamo:

$$Q_c = Q_{c-\text{modern}} \exp\left(-(t_{\text{Gyr}} - 4.5)/7\right). \quad (3)$$

The modern core flux is assumed to equal $Q_{c-\text{modern}} = 12$ TW because this is approximately the maximum core heat flow consistent with the absence of a dynamo assuming no inner core nucleation (Driscoll & Bercovici 2014; Nimmo 2015). We do not consider uncertainty in core heat flow because we are already incorporating a large factor of four range in radiogenic mantle heat production.

---

[5] https://github.com/joshuakt/Venus-evolution

**Table A1**
Uncertain Parameter Ranges Sampled in Nominal Venus Monte Carlo Calculations

| Parameter Type | Parameter | Nominal Range | References/Notes |
|---|---|---|---|
| Initial conditions | Water | $10^{19.5}$–$10^{21.5}$ kg | 2%–200% Earth oceans |
| | Carbon dioxide | $10^{20.4}$–$10^{21.5}$ kg | Approximately 40–600 bars (pending other atmospheric constituents) |
| | Radionuclide U and Th inventory (relative Earth) | 0.5–2.0 | Scalar multiplication of Earth's radionuclide inventories in Lebrun et al. (2013) |
| | Mantle free oxygen | $2 \times 10^{21}$–$6 \times 10^{21}$ (kg) | This ensures a post-solidification mantle redox around Quartz-Fayalite-Magnetite buffer |
| | K/U ratio in silicates | 6000–8440 | Average from Venera and Vega landers (Kaula 1999) |
| Solar evolution and escape parameters | Early Sun rotation rate (relative modern) | 1.8–45 | Implies XUV saturation time of 6–226 Myr (Tu et al. 2015) |
| | Escape efficiency at low XUV flux, $\varepsilon_{lowXUV}$ | 0.01–0.3 | See escape section in Krissansen-Totton et al. (2021) |
| | Transition parameter for cold-trap diffusion limited to XUV-limited escape, $\lambda_{tra}$ | $10^{-2}$–$10^{2}$ | See escape section in Krissansen-Totton et al. (2021) |
| | XUV energy that contributes to XUV escape above hydrodynamic threshold, $\zeta$ | 0%–100% | See escape section in Krissansen-Totton et al. (2021) |
| | Upper atmosphere temperature, $T_{meso}$ | 180–220 K | Samples modern Venus to Earth-like stratospheric temperatures (Mahieux et al. 2015; Pätzold et al. 2007) |
| Carbon cycle parameters | Temperature dependence of continental weathering, $T_{efold}$ | 5–30 K | Plausible Earth-like range (Krissansen-Totton et al. 2018) |
| | $CO_2$ dependence of continental weathering, $\gamma$ | 0.1–0.5 | Plausible Earth-like range (Krissansen-Totton et al. 2018) |
| | Weathering supply limit, $W_{sup-lim}$ | $10^{5}$–$10^{7}$ kg s$^{-1}$ | Broad terrestrial planet range (Foley 2015) |
| | Ocean calcium concentration, $[Ca^{2+}]$ | $10^{-4}$–0.3 mol kg$^{-1}$ | Plausible range for diverse terrestrial planet compositions (Kite & Ford 2018; Krissansen-Totton et al. 2018) |
| | Ocean carbonate saturation, $\Omega$ | 1–10 | (Zeebe & Westbroek 2003) |
| Interior evolution parameter | Solid-mantle viscosity coefficient, $V_{coef}$ | $10^{1}$–$10^{3}$ Pa s | Solid-mantle kinematic viscosity, $\nu_{rock}$ (m$^2$ s$^{-1}$), is given by the following equation: $\nu_{rock} = V_{coef} 3.8 \times 10^{7} \exp\left(\frac{350,000}{8.3147_p}\right)/\rho_m$; see Krissansen-Totton et al. (2021) |
| | Transition time from plate tectonics to stagnant lid tectonic regime, $t_{PT-SL}$ | 0.05–4 Gyr | See tectonics section in Appendix A.2; a transition anytime prior to 0.5 Ga is permitted |
| Crustal sinks oxygen and hydrological cycle parameters | Crustal hydration efficiency, fr$_{hydr-frac}$ | $10^{-3}$–0.03 | Upper limit wt % $H_2O$ in oceanic crust; lower limit hydration limited by cracking |
| | Dry oxidation efficiency, $f_{dry-oxid}$ | $10^{-4}$–10% | Plausible range of processes for Venus (Gillmann et al. 2009) |
| | Wet oxidation efficiency, $f_{wet-oxid}$ | $10^{-3}$–$10^{-1}$ | Based on oxidation of Earth's oceanic crust (Lécuyer & Ricard 1999) |
| | Maximum fractional molten area, $f_{lava}$ | $10^{-4}$–1.0 | See explanation in Krissansen-Totton et al. (2021) |
| | Max mantle water content, $M_{solid-H_2O-max}$ | 0.5–15 Earth oceans | Best-estimates maximum hydration of silicate mantle (Cowan & Abbot 2014) |
| Albedo parameters | Hot-state albedo, $A_H$ | 0.0–0.3 | See albedo parameterization in Krissansen-Totton et al. (2021) |
| | Cold-state albedo, $A_C$ | 0.2–0.7 | See albedo parameterization in Krissansen-Totton et al. (2021) |

During solid-state convection in a plate tectonics regime, thermal evolution is governed by the following equation:

$$\frac{4\pi\rho_m c_p (r_p{}^3 - r_c{}^3)}{3}\frac{dT_p}{dt}$$
$$= \frac{4\pi\rho_m q_r (r_p{}^3 - r_c{}^3)}{3} - 4\pi q_m r_p{}^2 + Q_c. \quad (4)$$

Finally, during solid-state convection in a stagnant lid regime, we also account for mantle cooling due to the advection of melt to the surface and subsequent cooling (Foley & Smye 2018):

$$\frac{4\pi\rho_m c_p (r_p{}^3 - r_c{}^3)}{3}\frac{dT_p}{dt} = \frac{4\pi\rho_m q_r (r_p{}^3 - r_c{}^3)}{3}$$
$$- 4\pi q_m r_p{}^2 + Q_c - MP\rho_{\text{melt}}(c_p(T_p - T_{\text{surf}}) + \text{H}_{\text{fusion}}). \quad (5)$$

The advective heat term depends on melt production, $MP$ ($\text{m}^3\,\text{s}^{-1}$), and the average melt density $\rho_{\text{melt}} = 3000\,\text{kg m}^{-3}$, which is assumed to be slightly less than that of the bulk mantle. Equations (2), (4), and (5) govern thermal evolution except in rare cases where transition to runaway greenhouse causes surface temperature to increase above mantle potential temperature, in which case a conduction regime is adopted (see Krissansen-Totton et al. 2021).

### A.2. Tectonic Models

Both plate tectonics and stagnant lid regimes are explicitly modeled. The plate tectonics model is identical to that described in Krissansen-Totton et al. (2021). Mantle heat flow, $q_m$ ($\text{W m}^{-2}$), is specified by the following parameterized convection equation:

$$q_m = \left(\frac{k}{d_{\text{convect}}}\right)(T_p - T_{\text{surf}})(\text{Ra}/\text{Ra}_{\text{crit}})^\beta. \quad (6)$$

Here $\text{Ra}_{\text{crit}} = 1100$ is the critical Rayleigh number, $k = 4.2$ $\text{W m}^{-1}\,\text{K}^{-1}$ is the thermal conductivity of silicates, $\beta = 1/3$, and $T_p$ and $T_{\text{surf}}$ are mantle potential temperature and mean surface temperature, respectively. The Rayleigh number, Ra, is given by

$$\text{Ra} = \frac{\alpha g (T_p - T_{\text{surf}}) d_{\text{convect}}{}^3}{\kappa\nu(T_p)}. \quad (7)$$

In this expression, $\alpha = 2 \times 10^{-5}\,\text{K}^{-1}$ is the thermal expansion coefficient for silicates, $g$ ($\text{m s}^{-12}$) is surface gravity, $\kappa = 10^{-6}$ $\text{m}^2\,\text{s}^{-1}$ is thermal diffusivity, and $\nu(T_p)$ is temperature-dependent kinematic mantle viscosity ($\text{m}^2\,\text{s}^{-1}$). Our parameterization for mantle viscosity—which is described in full in Krissansen-Totton et al. (2021)—ensures dynamic viscosities ranging from $10^{20}$ to $10^{22}$ Pa s for Earth-like mantle potential temperatures around 1600 K but smoothly transitions to magma ocean appropriate dynamic viscosities (<1 Pa s) above $\sim 1900$ K (see Krissansen-Totton et al. 2021, their Figure A3). The length scale of convection, $d_{\text{convect}}$, depends on whether a magma ocean is present:

$$d_{\text{convect}} = \begin{cases} r_p - r_s, & r_s < r_p \\ r_p - r_c, & r_s = r_p. \end{cases} \quad (8)$$

The radii refer to planetary radius, $r_p$, core radius, $r_c$, and solidification radius, $r_s$. The time evolution of the solidification radius is described in Krissansen-Totton et al. (2021).

For plate tectonics, we assume that the crust thickness, $\delta_{\text{crust}}$ (m), is equal to the thickness of the melt layer $d_{\text{melt}}$, which can be calculated from the average melt fraction, $\bar{\psi}$, and total melt volume:

$$d_{\text{melt}} = r_p - \left(r_p{}^3 - \frac{3}{4\pi}\bar{\psi}\int_{r=r(T=T_{\text{solidus}})}^{r=r_p} 4\pi r^2 dr\right)^{1/3}. \quad (9)$$

See Krissansen-Totton et al. (2021) for calculation of average melt fraction. Once crustal depth is known, the melt production rate, $MP$ ($\text{m}^3\,\text{s}^{-1}$), can be calculated from the plate velocity, $v_{\text{plate}}$ ($\text{m s}^{-1}$), and ridge length, $l_{\text{ridge}}$ (m):

$$MP = l_{\text{ridge}} \times d_{\text{melt}} \times v_{\text{plate}}. \quad (10)$$

From the expression for half-space cooling of oceanic crust (e.g., Turcotte & Schubert 2002 their Section 4.15; Kite et al. 2009), we can relate interior heat flow to ridge length and spreading rate as follows:

$$\left(\frac{q_m 4\pi r_p{}^2}{2k(T_p - T_{\text{surf}})}\right)^2 = \frac{4\pi r_p{}^2 \times l_{\text{ridge}} \times v_{\text{plate}}}{\pi\kappa}. \quad (11)$$

Substituting this into Equation (10), melt production can be calculated from only the heat flow and crustal depth:

$$MP = \left(\frac{q_m 4\pi r_p{}^2}{2k(T_p - T_{\text{surf}})}\right)^2 \frac{\pi\kappa}{4\pi r_p{}^2} d_{\text{melt}}. \quad (12)$$

Our stagnant lid model is based on that described in Foley & Smye (2018). In this framework, mantle heat flow, $q_m$, is specified by the following equation:

$$q_m = \frac{k}{2d_{\text{convect}}}(T_p - T_{\text{surf}})\theta^{-4/3}\text{Ra}^{1/3}. \quad (13)$$

Here the Frank–Kamenetskii parameter $\theta = 300{,}000 \times (T_p - T_{\text{surf}})/(8.314 \times T_p{}^2)$, whereas the internal Rayleigh number, Ra, is unchanged from plate tectonics regime. Melt production under the stagnant lid regime is calculated assuming cylindrical upwelling (Foley & Smye 2018):

$$MP = 17.8\pi r_p v_{\text{convect}}(d_{\text{melt}} - \delta)\bar{\psi}. \quad (14)$$

Here the convective velocity is assumed to be $v_{\text{convect}} = 0.05\kappa(\text{Ra}/\theta)^{2/3}/d_{\text{conv}}$ (Foley & Smye 2018), whereas the melting depth, $d_{\text{melt}}$, is now the depth at which mantle temperature intersects the solidus. This expression arises by considering the volumetric flow of upwelling melt out of a convective cylinder between the melting depth and the base of the lid. By analogy with Earth mantle convection, we assume the horizontal length of convective cells to be $0.45r_p$, which gives rise to the factor of 17.8 (Foley & Smye 2018). Average melt fraction, $\bar{\psi}$, is the volume-weighted melt fraction from the melting depth to the base of the stagnant lid, which has thickness, $\delta$ (m):

$$\bar{\psi} = \frac{\int_{r=r(T=T_{\text{solidus}})}^{r=r_p-\delta} \psi(T(r), T_{\text{solidus}}(P_{\text{overburden}}, r), T_{\text{liquidus}}(P_{\text{overburden}}, r)) \times 4\pi r^2 dr}{\int_{r=r(T=T_{\text{solidus}})}^{r=r_p-\delta} 4\pi r^2 dr}. \quad (15)$$

Here the melt fraction at any given radius is given by the following expression:

$$\psi(T(r), T_{\text{solidus}}, T_{\text{liquidus}}) = \begin{cases} 0, & T(r) < T_{\text{solidus}}(P_{\text{overburden}}, r) \\ 1, & T(r) > T_{\text{liquidus}}(P_{\text{overburden}}, r) \\ \frac{T(r) - T_{\text{solidus}}(P_{\text{overburden}}, r)}{T_{\text{liquidus}}(P_{\text{overburden}}, r) - T_{\text{solidus}}(P_{\text{overburden}}, r)}, & \text{otherwise} \end{cases}.$$

(16)

The expressions for the solidus and liquidus are described in Krissansen-Totton et al. (2021) but are derived from linear fits to the solidus for low-pressure dry peridotite and high-pressure lower mantle (Hirschmann 2000; Schaefer et al. 2016). Following Schaefer et al. (2016), the liquidus is assumed to be 600 K warmer than the solidus at all pressures. We also allow for modulation of the solidus and liquidus by the pressure overburden of surface volatiles. Here $P_{\text{overburden}}$ is the pressure from all $H_2O$ (liquid and gaseous), $CO_2$, and $O_2$ at the surface. There is a separate dependence on solid-body radius, $r$, to add pressure overburden from overlying silicate material.

The time evolution of lid thickness is given by the following system of equations (Foley & Smye 2018):

$$\frac{d\delta}{dt} = \frac{-q_m - k\frac{dT}{dr}\big|_{r=r_p-\delta}}{\rho_m c_p (T_p - T_l)}.$$

(17)

Here $\rho_m$ is mantle density and $c_p$ is mantle heat capacity. The lid base temperature, $T_l$, is specified by the following:

$$T_l = T_p - \frac{2.5 \times 8.314 \times T_p^2}{300,000}.$$

(18)

Here we have assumed an activation energy for mantle viscosity of 300 kJ mol$^{-1}$ (Karato & Wu 1993). The temperature gradient at the base of the lid is specified as follows:

$$k\frac{dT}{dr}\Big|_{r=r_p-\delta}$$
$$= \begin{cases} -k(T_l - T_c)/(\delta - \delta_{\text{crust}}) + 0.5 x_m(\delta - \delta_{\text{crust}}), & |T_c - T_l| > 1 \\ -k(T_l - T_{\text{surf}})/\delta + 0.5 x_c \delta_{\text{crust}}, & |T_c - T_l| \leqslant 1 \end{cases}$$

(19)

The terms $x_m$ and $x_c$ represent mantle and crustal radiogenic heat production:

$$x_m = Q_r \rho_m \varphi_{\text{mantle}} 4\pi(r_p^3 - r_c^3)/(3V_{\text{mantle}}),$$

(20)

$$x_c = Q_r \rho_m \varphi_{\text{crust}} 4\pi(r_p^3 - r_c^3)/(3V_{\text{crust}}).$$

(21)

The temperature at the base of the crust, $T_c$, is given by

$$T_c = \frac{T_{\text{surf}}(\delta - \delta_{\text{crust}}) + T_l \delta_{\text{crust}}}{\delta}$$
$$+ \frac{x_c \delta_{\text{crust}}^2(\delta - \delta_{\text{crust}}) + x_m \delta_{\text{crust}}(\delta - \delta_{\text{crust}})^2}{2k\delta}.$$

(22)

The terms $\varphi_{\text{mantle}}$ and $\varphi_{\text{crust}}$ denote the proportions of radionuclides partitioned into the mantle and crust, respectively. Initially, all radionuclides reside in the mantle, but once a stagnant lid begins to form, they evolve according to the following system of equations:

$$\frac{d\varphi_{\text{crust}}}{dt} = \frac{\varphi_{\text{mantle}} MP}{(V_{\text{mantle}} - V_{\text{crust}})} \frac{(1 - (1 - \bar{\psi})^{1/D_{\text{radiogenic}}})}{\bar{\psi}}$$
$$- \frac{\varphi_{\text{crust}}}{V_{\text{crust}}} \left(MP - 4\pi(r_p - \delta)^2 \min(0, \frac{d\delta}{dt})\right)(\tanh(20(\delta_{\text{crust}} - \delta)) + 1)$$

(23)

$$\frac{d\varphi_{\text{mantle}}}{dt} = -\frac{\varphi_{\text{mantle}} MP}{(V_{\text{mantle}} - V_{\text{crust}})} \frac{(1 - (1 - \bar{\psi})^{1/D_{\text{radiogenic}}})}{\bar{\psi}}$$
$$+ \frac{\varphi_{\text{crust}}}{V_{\text{crust}}} \left(MP - 4\pi(r_p - \delta)^2 \min(0, \frac{d\delta}{dt})\right)(\tanh(20(\delta_{\text{crust}} - \delta)) + 1).$$

(24)

Here $D_{\text{radiogenic}} = 0.002$ is the assumed average partition coefficient for radionuclides of U and Th (K is partitioned differently—see below). In the case where melt production is zero, the first term in both Equations (23) and (24) equals zero and the exchange of radiogenic material is governed entirely by the lid evolution.

The time evolution of crustal thickness is given by

$$\frac{dV_{\text{crust}}}{dt} = MP - \left(MP - 4\pi(r_p - \delta)^2 \times \min\left(0, \frac{d\delta}{dt}\right)\right)$$
$$\times (\tanh(20(\delta_{\text{crust}} - \delta)) + 1).$$

(25)

The depth of the crust is calculated from the crustal volume at each time step:

$$\delta_{\text{crust}} = r_p - \left(r_p^3 - \frac{3V_{\text{crust}}}{4\pi}\right)^{1/3}.$$

(26)

For full generality, we assume a transition from plate tectonics to stagnant lid anytime from 0.05 to 4 Gyr after magma ocean solidification (randomly sampled)

### A.3. Modeling $^{40}Ar$ Atmospheric Evolution

The $^{40}$Ar–$^{40}$K system is modeled by explicitly tracking the time evolution of $^{40}$K in the mantle, $^{40}$K$_{\text{mantle}}$, and lid, $^{40}$K$_{\text{lid}}$ (during stagnant lid), as well as its decay to mantle argon, $^{40}$Ar$_{\text{mantle}}$, and the accumulation of $^{40}$Ar in the atmosphere, $^{40}$Ar$_{\text{atmo}}$. Our approach follows that described in O'Rourke & Korenaga (2015). While plate tectonics is operating, the time evolution of each reservoir is given by the following system of equations:

$$\frac{d^{40}K_{\text{mantle}}}{dt} = -(\lambda_{\text{Ar}} + \lambda_{\text{Ca}})^{40}K_{\text{mantle}}/10^9 \, yr$$

(27)

$$\frac{d^{40}Ar_{\text{mantle}}}{dt} = \lambda_{\text{Ar}}{}^{40}K_{\text{mantle}}/10^9 \, yr - \rho_m MP \frac{^{40}Ar_{\text{mantle}}}{M_{\text{mantle}}}$$

(28)

$$\frac{d^{40}K_{\text{lid}}}{dt} = 0$$

(29)

$$\frac{d^{40}Ar_{\text{atmo}}}{dt} = \rho_m MP \frac{^{40}Ar_{\text{mantle}}}{M_{\text{mantle}}}.$$

(30)

Here $\lambda_{\text{Ar}} = 0.0581 \, \text{Gyr}^{-1}$ and $\lambda_{\text{Ca}} = 0.4862 \, \text{Gyr}^{-1}$ are decay constants for the decay of $^{40}$K to $^{40}$Ar and $^{40}$Ca, respectively, and $M_{\text{mantle}}$ is the mass of the mantle. Mantle $^{40}$Ar is transferred to the atmosphere via magmatic production.

When a stagnant lid is present, the time evolution of the K–Ar system is specified by the following system of equations:

$$
\begin{aligned}
\frac{d^{40}\text{K}_{\text{mantle}}}{dt} = & -(\lambda_{\text{Ar}} + \lambda_{\text{Ca}})\,{}^{40}\text{K}_{\text{mantle}}/10^9\,\text{yr} - \rho_m \\
& \times \frac{{}^{40}\text{K}_{\text{mantle}}MP}{M_{\text{mantle}}}\frac{(1 - (1 - \bar{\psi})^{1/D_K})}{\bar{\psi}} \\
& + \frac{{}^{40}\text{K}_{\text{lid}}}{V_{\text{crust}}}\left(MP - 4\pi(r_p - \delta)^2 \times \min\left(0, \frac{d\delta}{dt}\right)\right) \\
& \times (\tanh(20(\delta_{\text{crust}} - \delta)) + 1)
\end{aligned}
\tag{31}
$$

$$
\frac{d^{40}\text{Ar}_{\text{mantle}}}{dt} = \lambda_{\text{Ar}}\,{}^{40}\text{K}_{\text{mantle}}/10^9\,\text{yr} - \rho_m MP \frac{{}^{40}\text{Ar}_{\text{mantle}}}{M_{\text{mantle}}}
\tag{32}
$$

$$
\begin{aligned}
\frac{d^{40}\text{K}_{\text{lid}}}{dt} = & -(\lambda_{\text{Ar}} + \lambda_{\text{Ca}}) \\
& \times {}^{40}\text{K}_{\text{lid}}/10^9\,\text{yr} + \rho_m \frac{{}^{40}\text{K}_{\text{mantle}}MP}{M_{\text{mantle}}}\frac{(1 - (1 - \bar{\psi})^{1/D_K})}{\bar{\psi}} \\
& - \frac{{}^{40}\text{K}_{\text{lid}}}{V_{\text{crust}}}\left(MP - 4\pi(r_p - \delta)^2 \times \min\left(0, \frac{d\delta}{dt}\right)\right) \\
& \times (\tanh(20(\delta_{\text{crust}} - \delta)) + 1)
\end{aligned}
\tag{33}
$$

$$
\frac{d^{40}\text{Ar}_{\text{atmo}}}{dt} = \lambda_{\text{Ar}}\,{}^{40}\text{K}_{\text{lid}}/10^9\,\text{yr} + \rho_m MP \frac{{}^{40}\text{Ar}_{\text{mantle}}}{M_{\text{mantle}}}.
\tag{34}
$$

Here $D_K = 1$ is the assumed partition coefficient for potassium (O'Rourke & Korenaga 2015). Terms involving melt fraction equal zero when no melt is being produced. Mantle $^{40}$Ar is transferred to the atmosphere via magmatic production, whereas the decay of $^{40}$K in the lid is assumed to instantaneously add $^{40}$Ar to the atmosphere (O'Rourke & Korenaga 2015), which may result in somewhat overestimating atmospheric $^{40}$Ar, and perhaps explains why our median model estimate for modern $^{40}$Ar is larger than the observed atmospheric abundance (Figures 2(k) and 3(k)). We are additionally assuming that all argon in the mantle-derived source rock partitions into the melt. $^{40}$K in the mantle and lid are exchanged as the depth of the lithosphere evolves.

### A.4. Modeling $^4$He Atmospheric Evolution

A similar approach is adopted to calculate the accumulation of $^4$He, except in the case of helium, where it is generated via the decay of $^{235}$U, $^{238}$U, and $^{232}$Th, and so it is necessary to explicitly track the evolution of all three species in the mantle and the stagnant lid. For this, we follow the modeling framework of Namiki & Solomon (1998). While plate tectonics is operating, the governing equations are as follows:

$$
\frac{d^{238}\text{U}_{\text{mantle}}}{dt} = -\lambda_{238\text{U}}\,{}^{238}\text{U}_{\text{mantle}}; \quad \frac{d^{238}\text{U}_{\text{lid}}}{dt} = 0
\tag{35}
$$

$$
\frac{d^{235}\text{U}_{\text{mantle}}}{dt} = -\lambda_{235\text{U}}\,{}^{235}\text{U}_{\text{mantle}}; \quad \frac{d^{235}\text{U}_{\text{lid}}}{dt} = 0
\tag{36}
$$

$$
\frac{d^{232}\text{Th}_{\text{mantle}}}{dt} = -\lambda_{232\text{Th}}\,{}^{232}\text{Th}_{\text{mantle}}; \quad \frac{d^{232}\text{Th}_{\text{lid}}}{dt} = 0.
\tag{37}
$$

Here $^{238}$U$_{\text{mantle}}$, $^{235}$U$_{\text{mantle}}$, and $^{232}$Th$_{\text{mantle}}$ denote the mass of each radionuclide in the mantle, whereas $\lambda_{238\text{U}}^{238}$, $\lambda_{235\text{U}}$, and $\lambda_{232\text{Th}}$ are respective decay constants. The decay of these isotopes produces helium in the mantle, which can be transferred to the atmosphere via magmatic outgassing:

$$
\begin{aligned}
\frac{d^4\text{He}_{\text{mantle}}}{dt} = & \, 8\left(\frac{4}{238}\right)\lambda_{238\text{U}}\,{}^{238}\text{U}_{\text{mantle}} \\
& + 7\left(\frac{4}{235}\right)\lambda_{235\text{U}}\,{}^{235}\text{U}_{\text{mantle}} \\
& + 6\left(\frac{4}{232}\right)\lambda_{232\text{Th}}\,{}^{232}\text{Th}_{\text{mantle}} - \rho_m MP \frac{{}^4\text{He}_{\text{mantle}}}{M_{\text{mantle}}}
\end{aligned}
\tag{38}
$$

$$
\frac{d^4\text{He}_{\text{atmo}}}{dt} = \rho_m MP \frac{{}^4\text{He}_{\text{mantle}}}{M_{\text{mantle}}} - E_{\text{He}}\mu_{\text{He}}.
\tag{39}
$$

Here $^4$He$_{\text{mantle}}$ and $^4$He$_{\text{atmo}}$ denote the mass of 4-helium in the mantle and atmosphere, respectively. The contributions to mantle helium are weighted by their mass ratios and decay efficiencies. The mass flux of helium to space, $E_{\text{He}}\mu_{\text{He}}$, is described below.

During stagnant lid tectonics, the evolution of helium is governed by the following system of equations:

$$
\begin{aligned}
\frac{d^{238}\text{U}_{\text{mantle}}}{dt} = & -\lambda_{238\text{U}}\,{}^{238}\text{U}_{\text{mantle}} \\
& - \rho_m \frac{{}^{238}\text{U}_{\text{mantle}}MP}{M_{\text{mantle}}}\frac{(1 - (1 - \bar{\psi})^{1/D_{\text{radiogenic}}})}{\bar{\psi}} \\
& + \frac{{}^{238}\text{U}_{\text{lid}}}{V_{\text{crust}}}\left(MP - 4\pi(r_p - \delta)^2 \times \min(0, \frac{d\delta}{dt})\right) \\
& \times (\tanh(20(\delta_{\text{crust}} - \delta)) + 1)
\end{aligned}
\tag{40}
$$

$$
\begin{aligned}
\frac{d^{235}\text{U}_{\text{mantle}}}{dt} = & -\lambda_{235\text{U}}\,{}^{235}\text{U}_{\text{mantle}} \\
& - \rho_m \frac{{}^{235}\text{U}_{\text{mantle}}MP}{M_{\text{mantle}}}\frac{(1 - (1 - \bar{\psi})^{1/D_{\text{radiogenic}}})}{\bar{\psi}} \\
& + \frac{{}^{235}\text{U}_{\text{lid}}}{V_{\text{crust}}}\left(MP - 4\pi(r_p - \delta)^2 \times \min\left(0, \frac{d\delta}{dt}\right)\right) \\
& \times (\tanh(20(\delta_{\text{crust}} - \delta)) + 1)
\end{aligned}
\tag{41}
$$

$$
\begin{aligned}
\frac{d^{232}\text{Th}_{\text{mantle}}}{dt} = & -\lambda_{232\text{Th}}\,{}^{232}\text{Th}_{\text{mantle}} \\
& - \rho_m \frac{{}^{232}\text{Th}_{\text{mantle}}MP}{M_{\text{mantle}}}\frac{(1 - (1 - \bar{\psi})^{1/D_{\text{radiogenic}}})}{\bar{\psi}} \\
& + \frac{{}^{232}\text{Th}_{\text{lid}}}{V_{\text{crust}}}\left(MP - 4\pi(r_p - \delta)^2 \times \min\left(0, \frac{d\delta}{dt}\right)\right) \\
& \times (\tanh(20(\delta_{\text{crust}} - \delta)) + 1)
\end{aligned}
\tag{42}
$$

$$
\begin{aligned}
\frac{d^{238}\text{U}_{\text{lid}}}{dt} = & -\lambda_{238\text{U}}\,{}^{238}\text{U}_{\text{lid}} \\
& + \rho_m \frac{{}^{238}\text{U}_{\text{lid}}MP}{M_{\text{mantle}}}\frac{(1 - (1 - \bar{\psi})^{1/D_{\text{radiogenic}}})}{\bar{\psi}} \\
& - \frac{{}^{238}\text{U}_{\text{lid}}}{V_{\text{crust}}}\left(MP - 4\pi(r_p - \delta)^2 \times \min\left(0, \frac{d\delta}{dt}\right)\right) \\
& \times (\tanh(20(\delta_{\text{crust}} - \delta)) + 1)
\end{aligned}
\tag{43}
$$

$$\frac{d^{235}U_{lid}}{dt} = -\lambda_{235U}\,^{235}U_{lid}$$
$$+ \rho_m \frac{^{235}U_{lid}MP}{M_{mantle}} \frac{(1 - (1 - \bar{\psi})^{1/D_{radiogenic}})}{\bar{\psi}}$$
$$- \frac{^{235}U_{lid}}{V_{crust}}\left(MP - 4\pi(r_p - \delta)^2 \times \min\left(0, \frac{d\delta}{dt}\right)\right)$$
$$\times (\tanh(20(\delta_{crust} - \delta)) + 1) \tag{44}$$

$$\frac{d^{232}Th_{lid}}{dt} = -\lambda_{232Th}\,^{232}Th_{lid}$$
$$+ \rho_m \frac{^{232}Th_{lid}MP}{M_{mantle}} \frac{(1 - (1 - \bar{\psi})^{1/D_{radiogenic}})}{\bar{\psi}}$$
$$- \frac{^{232}Th_{lid}}{V_{crust}}\left(MP - 4\pi(r_p - \delta)^2 \times \min(0, \frac{d\delta}{dt})\right)$$
$$\times (\tanh(20(\delta_{crust} - \delta)) + 1) \tag{45}$$

$$\frac{d^4He_{mantle}}{dt} = 8\left(\frac{4}{238}\right)\lambda_{238U}\,^{238}U_{mantle}$$
$$\times + 7\left(\frac{4}{235}\right)\lambda_{235U}\,^{235}U_{mantle} + 6\left(\frac{4}{232}\right)$$
$$\times \lambda_{232Th}\,^{232}Th_{mantle} - \rho_m MP \frac{^4He_{mantle}}{M_{mantle}} \tag{46}$$

$$\frac{d^4He_{atmo}}{dt} = \rho_m MP \frac{^4He_{mantle}}{M_{mantle}}$$
$$+ 8\left(\frac{4}{238}\right)\lambda_{238U}\,^{238}U_{lid} + 7\left(\frac{4}{235}\right)$$
$$\times \lambda_{235U}\,^{235}U_{lid} + 6\left(\frac{4}{232}\right)\lambda_{232Th}\,^{232}Th_{lid} - E_{He}\mu_{He}. \tag{47}$$

The escape flux of He during all tectonic modes is approximated using the same parameterization, with first-order dependence on the solar XUV flux and atmospheric abundance (Namiki & Solomon [1998]). The flux is scaled to match the observed modern escape rate of helium of $9.6 \times 10^5$ kg yr$^{-1}$ (Prather & McElroy [1983]; Krasnopolsky & Gladstone [2005]):

$$E_{He}\mu_{He} = 9.6 \times 10^5 \left(\frac{F_{XUV}(t)}{F_{XUV}(4.5\,Gyr)}\right)\left(\frac{f_{He}}{10\,ppm}\right). \tag{48}$$

The modern helium abundance assumed for this scaling law is taken to be 10 ppm (Krasnopolsky & Gladstone [2005]). However, it should be noted that this is an extrapolation to the mid-lower atmosphere, and the dependence of helium escape fluxes on XUV fluxes is likely more complex and composition dependent than Equation ([48]) suggests. Our helium calculations should be taken as crude approximations to Venus's atmospheric helium abundance evolution.

### A.5. Stagnant Lid Volatile Cycling

The cycling of oxygen, water, and carbon dioxide between the atmosphere and the mantle is described in Krissansen-Totton et al. ([2021]). The following changes were made to accommodate volatile cycling under a stagnant lid regime.

The water content of the stagnant lid, $M_{Lid-H_2O}$, is explicitly calculated using the following equation:

$$\frac{dM_{Lid-H_2O}}{dt} = fr_{hydr-frac}f_{hydr-depth}MP \times \min$$
$$\times \left\{\frac{d_{ocean}}{d_{max-ocean}}, 1.0\right\} \times \max\left\{0, 1 - \frac{M_{Lid-H_2O}}{M_{lid-H_2O-max}}\right\}. \tag{49}$$

Here $fr_{hydr-frac} = 10^{-3}$–0.03 (sampled uniformly in log space) is the unknown efficiency of hydration reactions. We are assuming that, at most, hydrated crust is 3% water by mass (Schaefer & Sasselov [2015]). We assume a linear dependence on ocean depth for as long as there is emerged land, then no ocean depth dependence beyond this. Additionally, we assume that the return of water to the interior tapers off as the water content of the lid approaches its maximum value, $M_{lid-H_2O-max}$ (kg). The fractional depth to which hydration occurs is given by the following expressions:

$$d_{hydr} = \max\left\{0, \delta_{crust}\left(\frac{973 - T_{surf}}{T_p - T_{surf}}\right)\right\} \tag{50}$$

$$f_{hydr-depth} = \min\{d_{hydr}/\delta_{crust}, 1.0\}. \tag{51}$$

Here 973 K is the maximum surface temperature for serpentine stability (Schaefer & Sasselov [2015]). Since we assume that no hydration occurs below the crust, it is also helpful to define the fractional depth of hydration as the ratio of the hydration depth to the crustal depth, or 1.0 (whichever is smaller). The maximum water content of the lid is given by the following expression:

$$M_{lid-H_2O-max} = 0.03\rho_m V_{crust}f_{hydr-depth}. \tag{52}$$

The rate of lid hydration controls the rate of "wet" crustal oxidation via serpentinization reactions:

$$F_{H_2O-serp} = f_{wet-oxid}\left(\frac{dM_{Lid-H_2O}}{dt} \Big/ fr_{hydr-frac}\right)$$
$$\times fr_{Fe}\left(\frac{M_{solid-FeO}}{M_{solid-FeO} + M_{solid-FeO_{1.5}}}\right). \tag{53}$$

Here $f_{wet-oxid}$ is another unknown efficiency parameter ($10^{-3}$–$10^{-1}$) representing the fraction of crustal iron that is oxidized via hydration reactions (Lécuyer & Ricard [1999]), $fr_{Fe}$ is the total mass fraction of all iron-bearing species in the mantle (constant), and $M_{solid-FeO}$ and $M_{solid-FeO_{1.5}}$ represent the mass of ferrous and ferric iron in the mantle, respectively.

The loss of water from the surface and the gain of water in the silicate interior from hydration are given by the following expressions:

$$F_{ingas-H_2O-loss} = \frac{dM_{Lid-H_2O}}{dt} + F_{H_2O-serp}\frac{\mu_{H_2O}}{3\mu_{FeO}}$$
$$F_{ingas-H_2O-gain} = \frac{dM_{Lid-H_2O}}{dt}. \tag{54}$$

Here $\mu_{H_2O}$ and $\mu_{FeO}$ are the mean molecular weights of water and ferrous iron, respectively. Note that water loss via serpentinization does not add water to the interior because the reaction yields molecular hydrogen that is assumed to be lost to space via diffusion-limited escape (Krissansen-Totton et al. 2021).

To model carbon cycling under a stagnant lid, we adopted the same weathering model as in Krissansen-Totton et al. (2021), except that during stagnant lid only 10% of melt production is assumed to be available for seafloor weathering.

### A.6. Escape Parameterizations

The atmospheric escape parameterizations adopted in Krissansen-Totton et al. (2021) were chosen to minimize nonbiological oxygen accumulation. In the diffusion limit, H escape was limited by the diffusion of water through a noncondensible background gas (Wordsworth & Pierrehumbert 2013). In this study, we instead assume that eddy diffusion dominates vertical transport at altitudes where water is more abundant than atomic H and O (Catling & Kasting 2017), and so escape of hydrogen is limited by the diffusion of atomic hydrogen through the background atmosphere:

$$\varphi_{\text{diff}} = b_{\text{H}} f_{\text{H}} (1/\text{H}_n - 1/\text{H}_{\text{H}})$$
$$b_{\text{H}} = \frac{b_{\text{H}-CO_2}pCO_2 + b_{\text{H}-N_2}pN_2 + b_{\text{H}-O}pO}{pCO_2 + pN_2 + pO}$$
$$\text{H}_{\text{H}} = \frac{8.314T_{\text{meso}}}{\mu_{\text{H}} g}$$
$$\text{H}_n = \frac{8.314T_{\text{meso}}}{\bar{\mu} g}. \quad (55)$$

Here $b_{i-j}$ (mol m$^{-1}$ s$^{-1}$) is the binary diffusion coefficient of the $i$th species through the $j$th species (Marrero & Mason 1972; Zahnle & Kasting 1986). These are weighted by the stratospheric mixing ratios of each noncondensible constituent (CO$_2$, N$_2$, and atomic O), which are obtained from the atmospheric profile calculations and from assuming that all water is dissociated. The scale heights of hydrogen, H$_{\text{H}}$ (m), and of the background gases, H$_n$ (m), depend on upper atmosphere temperature, $T_{\text{meso}}$ (see main text). The hydrogen mixing ratio in the upper atmosphere, $f_{\text{H}}$, is assumed to be double the upper atmosphere water mixing ratio, $f_{\text{H}_2\text{O}}$. The diffusion-limited escape flux is $\varphi_{\text{diff}}$ (mol H m$^{-2}$ s$^{-1}$). Note that this diffusion limit only applies in the low H abundance regime and that escape is XUV limited for H-rich upper atmospheres (Krissansen-Totton et al. 2021).

### Appendix B
### Sensitivity Tests

#### B.1. Sensitivity of Results to Ion Escape and Melt Depletion

The ion escape sensitivity test described in the main text is shown in Figure B1. In our nominal model, the silicate solidus and liquidus depend on pressure but are independent of composition. This is reasonable under plate tectonics, whereby oceanic crust is assumed to be completely recycled into the mantle. However, under a stagnant lid regime, melt accumulation in the lid may progressively deplete the mantle, resulting in an increasing solidus. Melt production, and therefore oxygen sinks, may subsequently decline over time. To test the impact of mantle depletion, we repeated our calculations allowing for the evolution of the solidus with melt extraction from the mantle. During plate tectonics (or magma ocean), the solidus was assumed to remain constant, but under stagnant lid the solidus evolves as follows:

$$\frac{dT_{Solidus}}{dt} = \frac{MP \times (1 - \bar{\psi})}{V_{mantle}}(T_{Liquidus} - T_{Solidus}). \quad (56)$$

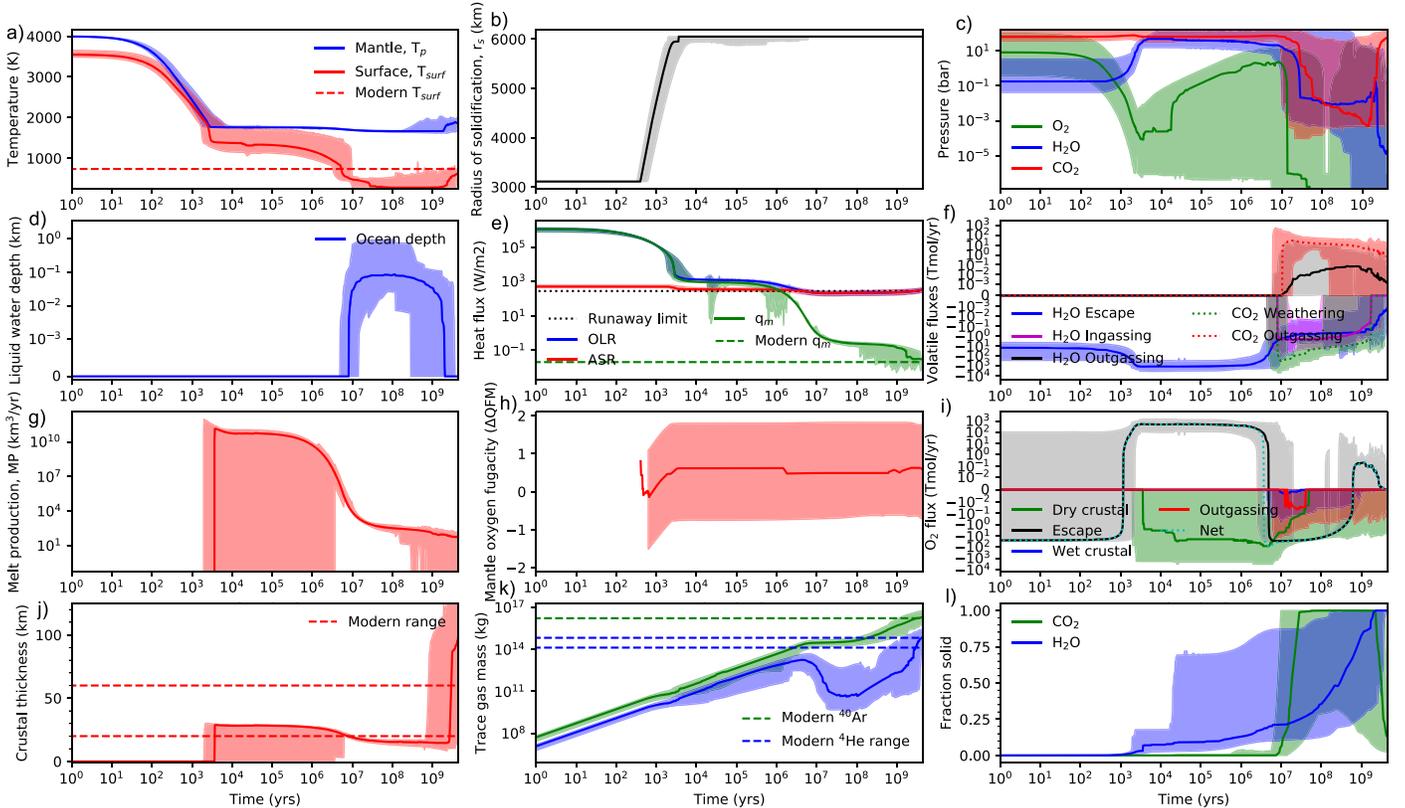

**Figure B1.** All habitable past model outputs from a sensitivity test where we allow for nonthermal escape of O$^+$ ions using estimates from Kulikov et al. (2006) (case 2b). Model outputs are largely the same as those in Figure 3 in the main text. Note that O$^+$ loss is potentially very large within the first <1 Gyr but is negligible subsequently, and so it does not affect late oxygen drawdown and the duration of habitability.

Here $MP$ is melt production as defined by Equation (14), $V_{mantle}$ is the volume of the mantle, $\bar{\psi}$ is the mean melt fraction as defined by Equation (15), and the liquidus is assumed to remain constant. Results from including mantle depletion are shown in Figures B2 and B3 and described in the main text.

### B.2. Sensitivity of Results to Volatile Retention in Mantle and Initial Mantle Redox

To investigate the sensitivity of our results to greater mantle volatile retention during magma ocean solidification, we repeated nominal calculations accounting for melt compaction in the freezing front. Following Hier-Majumder & Hirschmann (2017), we used the following expression to determine the melt fraction trapped in the mantle as the solidification front moves toward the surface, $f_{TL}$:

$$f_{TL} = -\frac{0.3\tau_c}{T_{Liquidus} - T_{Solidus}}\frac{dT_P}{dt}. \quad (57)$$

Here $\tau_c = 10^6$ yr is the assumed compaction timescale, which is derived from a $\sim$cm yr$^{-1}$ characteristic velocity of matrix sedimentation and $\sim$10 km freezing front (Hier-Majumder & Hirschmann 2017). Rapid magma ocean solidification (i.e., a large rate of change in mantle potential temperature) results in greater volatile retention. Additionally, we require that $f_{TL}$ is bounded below by 0 (negative trapped melt fractions are unphysical) and above by 0.3 (retained melt fraction does not exceed disaggregation melt fraction). The trapped melt fraction is incorporated into the system of equations governing water and carbon dioxide reservoir evolution by modifying Equation (S16) in Krissansen-Totton et al. (2021):

$$\frac{dM_{Solid-H_2O}}{dt} = 4\pi\rho_m f r_{H_2O} r_s^2 \frac{dr_s}{dt}(k_{H_2O}(1 - f_{TL}) + f_{TL})$$
$$+ F_{ingas-H_2O-gain} - F_{outgas-H_2O}$$

$$\frac{dM_{Fluid-H_2O}}{dt} = -4\pi\rho_m f r_{H_2O} r_s^2 \frac{dr_s}{dt}(k_{H_2O}(1 - f_{TL}) + f_{TL})$$
$$- F_{ingas-H_2O-loss} + F_{outgas-H_2O} - 0.5E_H\mu_{H_2O}$$

$$\frac{dM_{Solid-CO_2}}{dt} = 4\pi\rho_m f r_{CO_2} r_s^2 \frac{dr_s}{dt}(k_{CO_2}(1 - f_{TL}) + f_{TL})$$
$$+ F_{Weather-CO_2} - F_{outgas-CO_2}$$

$$\frac{dM_{Fluid-CO_2}}{dt} = -4\pi\rho_m f r_{CO_2} r_s^2 \frac{dr_s}{dt}(k_{CO_2}(1 - f_{TL}) + f_{TL})$$
$$- E_{CO_2}\mu_{CO_2} - F_{Weather-CO_2} + F_{outgas-CO_2}.$$

$$(58)$$

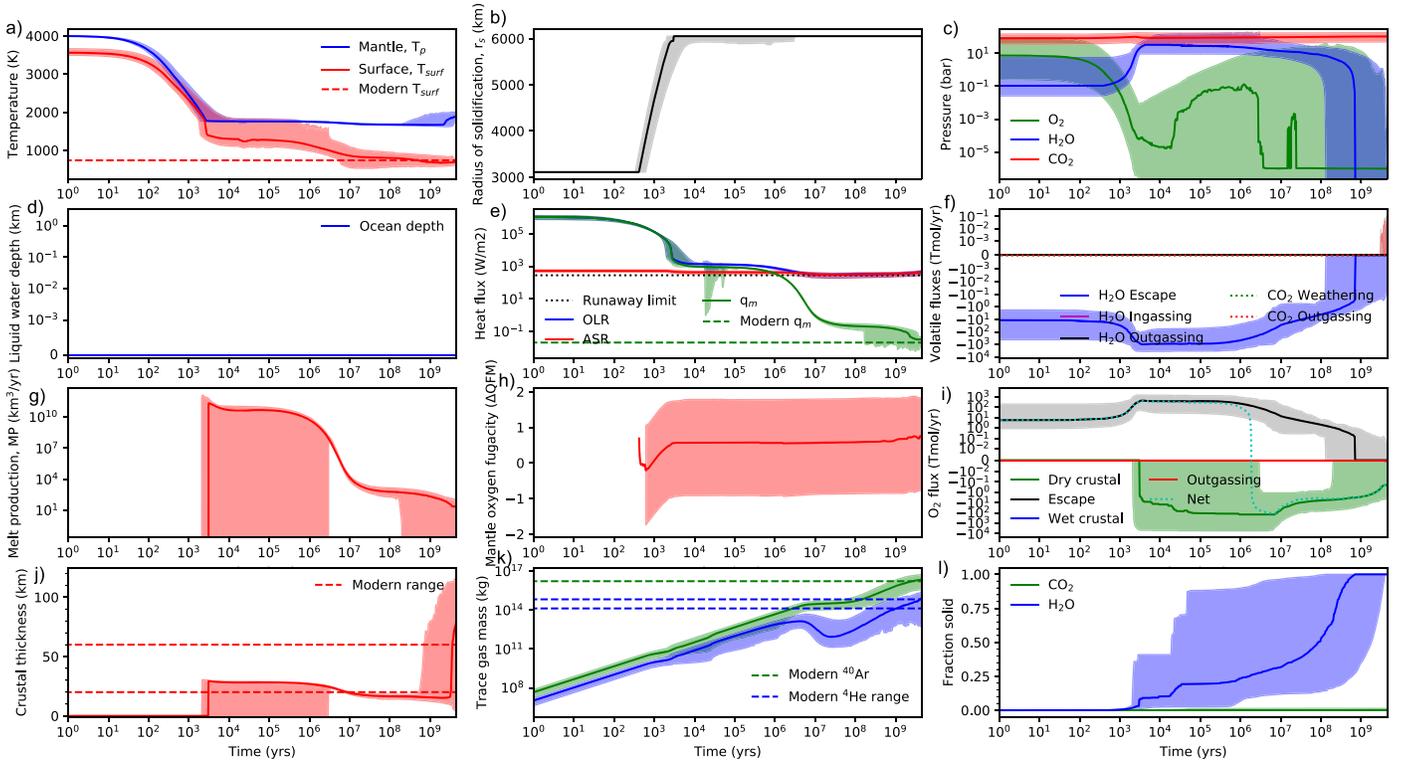

**Figure B2.** Sensitivity test accounting for increasing solidus due to mantle depletion and melt accumulation in the lid. All never-habitable model outputs that recover the modern atmosphere are plotted. Outcomes are comparable to those in Figure 2, except that average modern melt production (panel (g)) is slightly less than the nominal model.

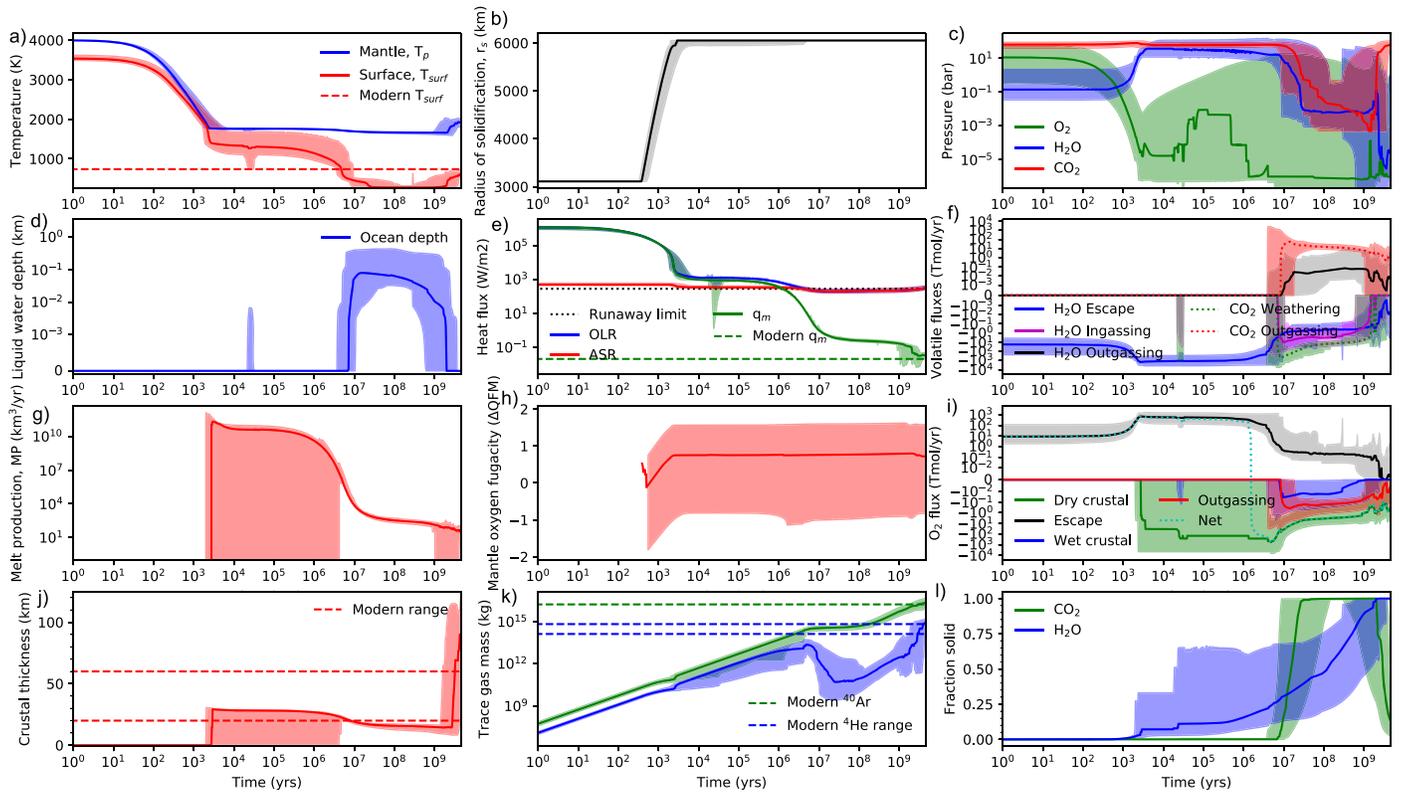

**Figure B3.** Sensitivity test accounting for increasing solidus due to mantle depletion and melt accumulation in the lid. All transiently habitable model outputs that recover the modern atmosphere are plotted. Outcomes are comparable to those in Figure 3, except that average modern melt production (panel (g)) is slightly less than the nominal model.

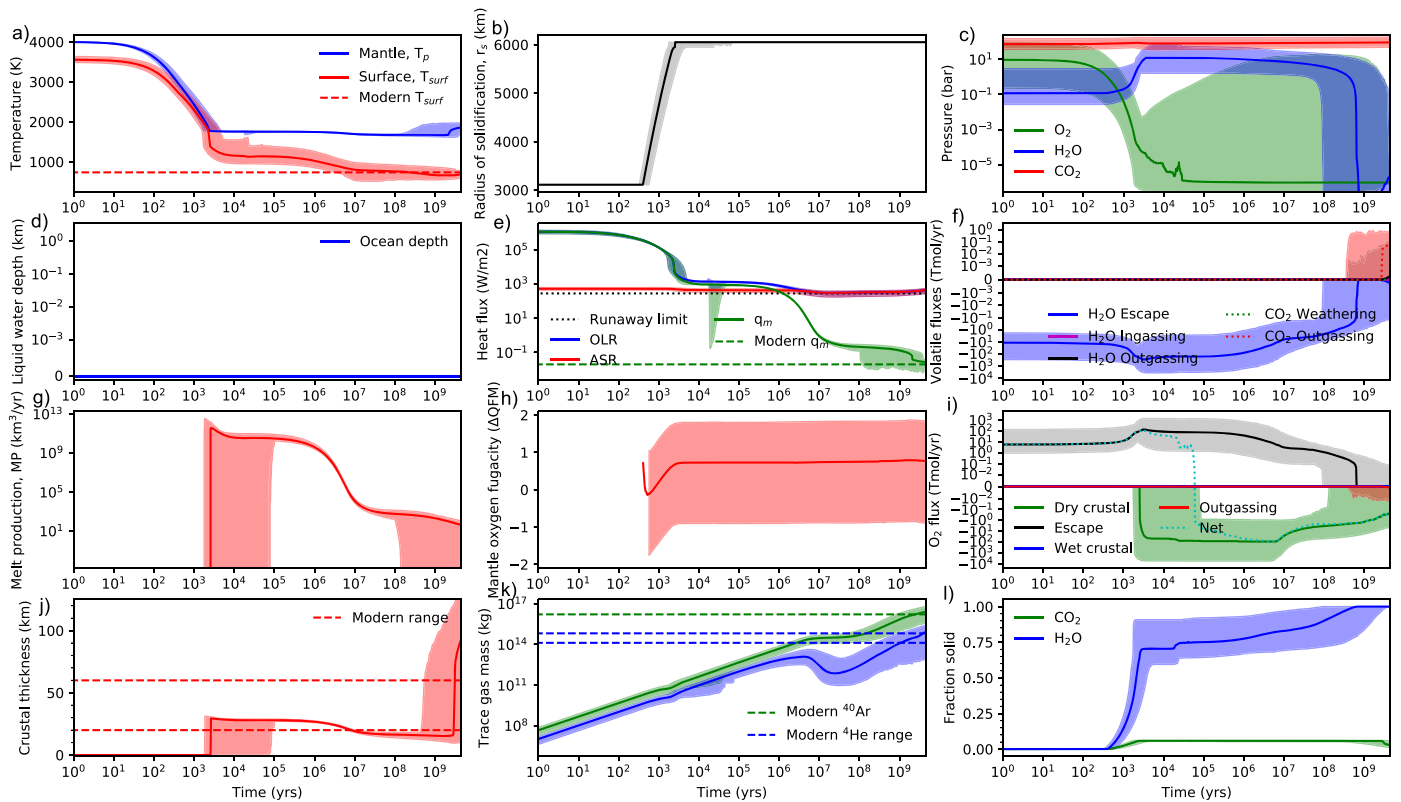

**Figure B4.** Sensitivity test with greater volatile retention in mantle during magma ocean solidification (panel (l)). All never-habitable model outputs that recover the modern atmosphere are plotted. Outcomes are comparable to those in Figure 2, except that some outgassing occurs throughout Venus's evolution (panels (f) and (i)).

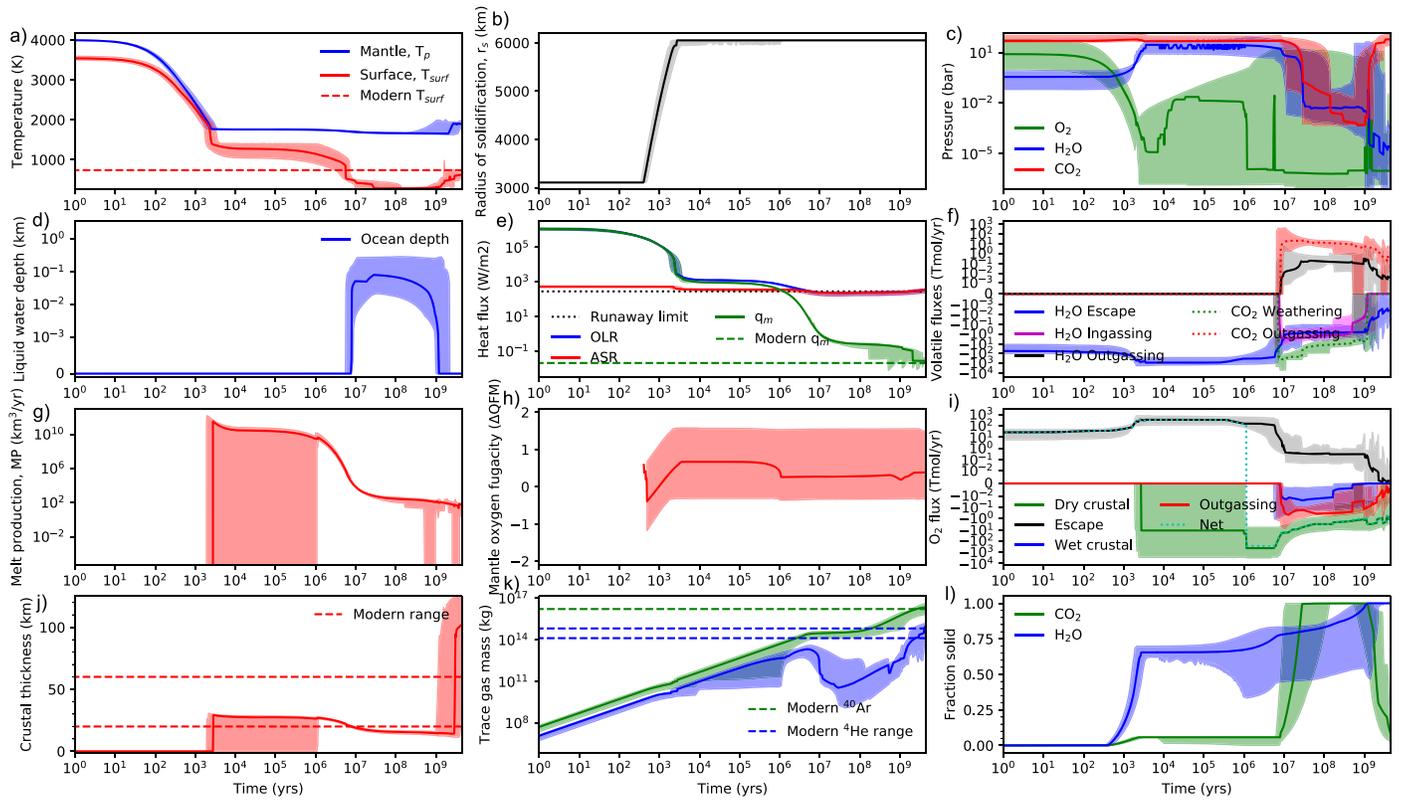

**Figure B5.** Sensitivity test with greater volatile retention in mantle during magma ocean solidification (panel (l)). All transiently habitable model outputs that recover the modern atmosphere are plotted. Outcomes are comparable to those in Figure 3, except that outgassing occurs throughout Venus's evolution (panels (f) and (i)), and the overall number of successful habitable model outputs is lower.

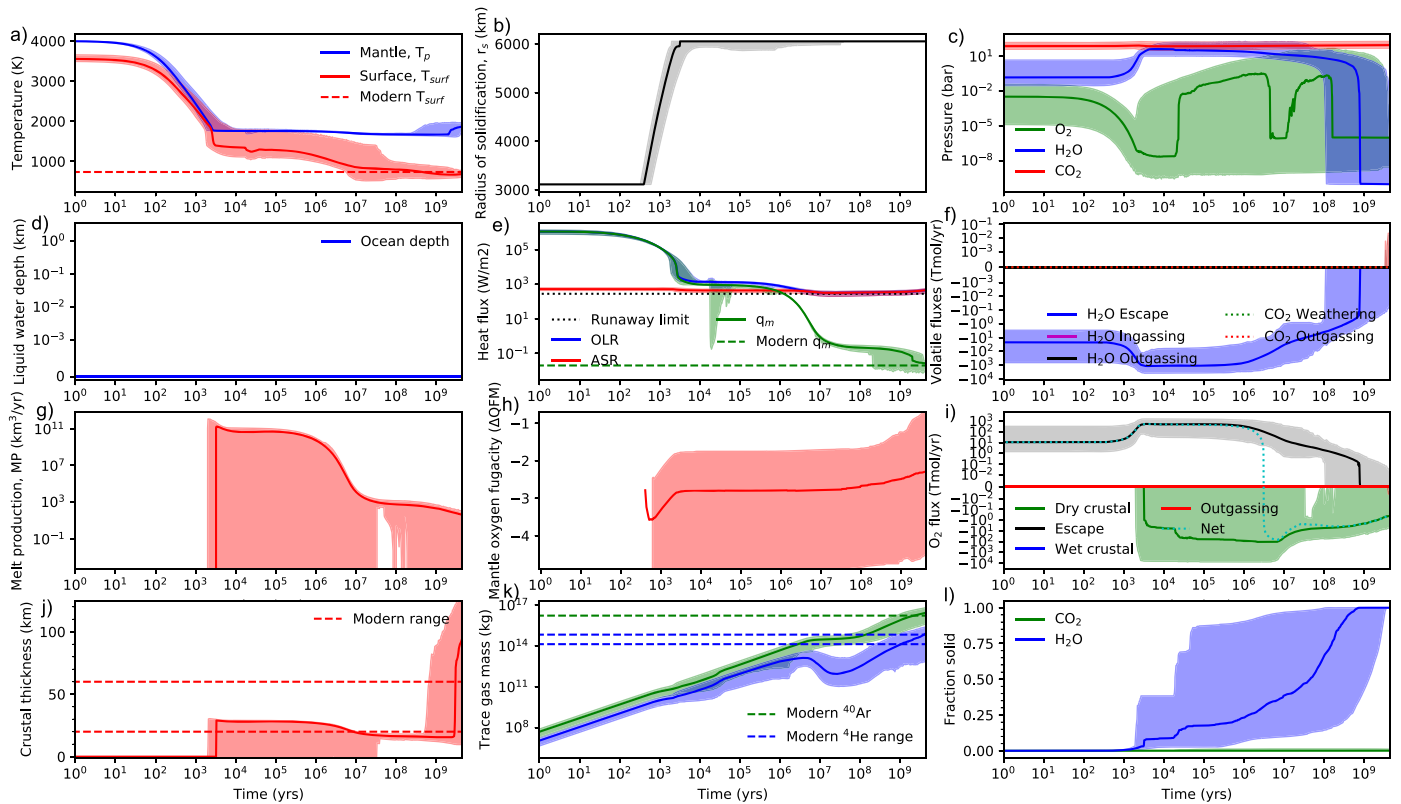

**Figure B6.** Sensitivity test with a more reducing initial mantle (panel (h)). All never-habitable model outputs that recover the modern atmosphere are plotted. Outcomes are comparable to those in Figure 2, except that the modern mantle redox state is closer to the IW buffer.

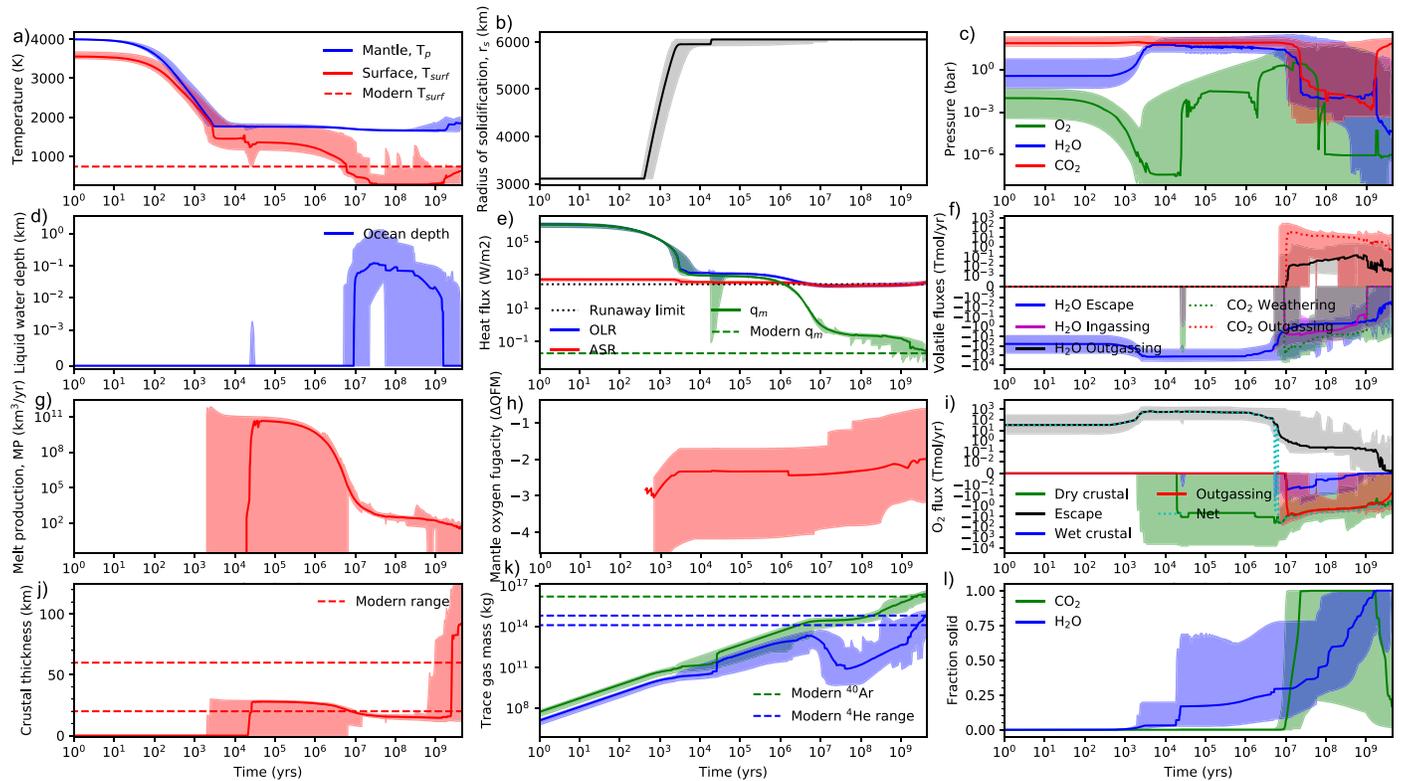

**Figure B7.** Sensitivity test with a more reducing initial mantle (panel (h)). All transiently habitable model outputs that recover the modern atmosphere are plotted. Outcomes are comparable to those in Figure 3, except that the modern mantle redox state is closer to the IW buffer.

The terms and parameter values in these equations are defined in detail in Krissansen-Totton et al. (2021). Broadly speaking, both a trapped melt fraction ($f_{TL}$) and solid-melt equilibrium partitioning ($k_{H_2O}$, $k_{CO_2}$) contribute to volatile retention in the growing solid silicate interior. Results from including volatile retention are shown in Figures B4 and B5 and described in the main text.

The initial mantle redox sensitivity tests described in the main text are shown in Figures B6 and B7.

## ORCID iDs

Jonathan J. Fortney ● https://orcid.org/0000-0002-9843-4354
Francis Nimmo ● https://orcid.org/0000-0003-3573-5915

## References

Akins, A. B., Lincowski, A. P., Meadows, V. S., & Steffes, P. G. 2021, ApJL, 907, L27
Armstrong, K., Frost, D. J., McCammon, C. A., Rubie, D. C., & Boffa Ballaran, T. 2019, Sci, 365, 903
Arney, G., Meadows, V., Crisp, D., et al. 2014, JGRE, 119, 1860
Arvidson, R. E., Greeley, R., Main, M. C., et al. 1992, JGR, 97, 13303
Baines, K. H., Atreya, S. K., Bullock, M. A., et al. 2013, in Comparative Climatology of Terrestrial Planets, ed. S J. Mackwell et al., Vol. 1 (Tucson, AZ: Univ. Arizona Press), 137
Bains, W., Petkowski, J. J., Seager, S., et al. 2021, AsBio, 21, 1277
Barnes, R., Mullins, K., Goldblatt, C., et al. 2013, AsBio, 13, 225
Barstow, J. K., Aigrain, S., Irwin, P. G. J., Kendrew, S., & Fletcher, L. N. 2016, MNRAS, 458, 2657
Berger, G., Cathala, A., Fabre, S., et al. 2019, Icar, 329, 8
Bézard, B., Fedorova, A., Bertaux, J.-L., Rodin, A., & Korablev, O. 2011, Icar, 216, 173
Bierson, C., & Zhang, X. 2020, JGRE, 125, e06159
Bjonnes, E., Hansen, V. L., James, B., & Swenson, J. B. 2012, Icar, 217, 451

Boukrouche, R., Lichtenberg, T., & Pierrehumbert, R. T. 2021, ApJ, 919, 130
Byrne, P. K., Ghail, R. C., Gilmore, M. S., et al. 2021, Geo, 49, 81
Catling, D. C., & Kasting, J. F. 2017, Atmospheric Evolution on Inhabited and Lifeless Worlds (Cambridge: Cambridge Univ. Press)
Catling, D. C., Zahnle, K. J., & McKay, C. P. 2001, Sci, 293, 839
Chassefière, E. 1996, Icar, 124, 537
Chassefière, E., Bertaux, J. L., Kurt, V. G., & Smirnov, A. S. 1986, P&SS, 34, 585
Chassefière, E., Wieler, R., Marty, B., & Leblanc, F. 2012, P&SS, 63, 15
Clough, S., Shephard, M. W., Mlawer, E. J., et al. 2005, JQSRT, 91, 233
Cowan, N. B., & Abbot, D. S. 2014, ApJ, 781, 27
Donahue, T., Hoffman, J. H., Hodges, R. R., & Watson, A. J. 1982, Sci, 216, 630
Driscoll, P., & Bercovici, D. 2013, Icar, 226, 1447
Driscoll, P., & Bercovici, D. 2014, PEPI, 236, 36
Ehrenreich, D., Vidal-Madjar, A., Widemann, T., et al. 2012, A&A, 537, L2
Elkins-Tanton, L. T. 2008, E&PSL, 271, 181
Encrenaz, T., Greathouse, T. K., Marq, E., et al. 2020, A&A, 643, L5
Eymet, V., Coustet, C., & Piaud, B. 2016, JPhCS, 676, 012005
Fedorov, A., Barabash, S., Sauvaud, J.-A., et al. 2011, JGRA, 116, A07220
Fegley, B., Zolotov, M. Y., & Lodders, K. 1997, Icar, 125, 416
Foley, B. J. 2015, ApJ, 812, 36
Foley, B. J., & Smye, A. J. 2018, AsBio, 18, 873
Garvin, J., Getty, S., Arney, G. N., et al. 2020, AGUFM, P026-0001
Gillmann, C., Chassefière, E., & Lognonné, P. 2009, E&PSL, 286, 503
Gillmann, C., Golabek, G. J., Raymond, S. N., et al. 2020, NatGe, 13, 265
Gillmann, C., & Tackley, P. 2014, JGRE, 119, 1189
Gilmore, M. S., Mueller, N., & Helbert, J. 2015, Icar, 254, 350
Gilmore, M., Treiman, A., Helbert, J., & Smrekar, S. 2017, SSRv, 212, 1511
Goldblatt, C., Robinson, T. D., Zahnle, K. J., & Crisp, D. 2013, NatGe, 6, 661
Greaves, J. S., Richards, A. M. S., Bains, W., et al. 2021, NatAs, 5, 655
Grinspoon, D. H. 1987, Sci, 238, 1702
Grinspoon, D. H. 1993, Natur, 363, 428
Hamano, K., Abe, Y., & Genda, H. 2013, Natur, 497, 607
Hansen, V. L., & López, I. 2010, Geo, 38, 311
Head, J. W., Campbell, D. B., Elachi, C., et al. 1991, Sci, 252, 276
Herrick, R. R., & Rumpf, M. E. 2011, JGRE, 116, E02004
Hier-Majumder, S., & Hirschmann, M. M. 2017, GGG, 18, 3078
Hirschmann, M. M. 2000, GGG, 1, 1042

Innanen, K., Mikkola, S., & Wiegert, P. 1998, AJ, 116, 2055
Johnson, B. W., & Goldblatt, C. 2018, GGG, 19, 2516
Johnstone, C. P., Güdel, M., Lammer, H., & Kislyakova, K. G. 2018, A&A, 617, A107
Kane, S. R., Arney, G., Crisp, D., et al. 2019, JGRE, 124, 2015
Kane, S. R., Vervoort, J., Horner, J., & Pozuelos, F. 2020, PSJ, 1, 42
Karato, S.-I., & Wu, P. 1993, Sci, 260, 771
Kasting, J. F. 1988, Icar, 74, 472
Kasting, J. F., & Pollack, J. B. 1983, Icar, 53, 479
Kaula, W. M. 1999, Icar, 139, 32
Khawja, S., Ernst, R. E., Samson, C., et al. 2020, NatCo, 11, 5789
Kite, E. S., & Ford, E. B. 2018, ApJ, 864, 75
Kite, E. S., Manga, M., & Gaidos, E. 2009, ApJ, 700, 1732
Kleinböhl, A., Willacy, K., Friedson, A. J., Chen, P., & Swain, M. R. 2018, ApJ, 862, 92
Krasnopolsky, V. A., & Gladstone, G. R. 2005, Icar, 176, 395
Krissansen-Totton, J., Arney, G. N., & Catling, D. C. 2018, PNAS, 115, 4105
Krissansen-Totton, J., Fortney, J. J., Nimmo, F., & Wogan, N. 2021, AGUA, 2, e00294
Kulikov, Y. N., Lammer, H., Lichtenegger, H. I. M., et al. 2006, P&SS, 54, 1425
Lammer, H., Leitzinger, M., Scherf, M., et al. 2020, Icar, 339, 113551
Lammer, H., Sproß, L., Grenfell, J. L., et al. 2019, AsBio, 19, 927
Lammer, H., Zerkle, A. L., Gebauer, S., et al. 2018, A&ARv, 26, 2
Lebrun, T., Massol, H., Chassefière, E., et al. 2013, JGRE, 118, 1155
Lécuyer, C., & Ricard, Y. 1999, E&PSL, 165, 197
Lincowski, A. P., Lustig-Yaeger, J., & Meadows, V. S. 2019, AJ, 158, 26
Lustig-Yaeger, J., Meadows, V. S., & Lincowski, A. P. 2019a, AJ, 158, 27
Lustig-Yaeger, J., Meadows, V. S., & Lincowski, A. P. 2019b, ApJL, 887, L11
Ma, Q., & Tipping, R. 1992, JChPh, 97, 818
Mahieux, A., Vandaele, A. C., Bougher, S. W., et al. 2015, P&SS, 113, 309
Marcq, E., Mills, F. P., Parkinson, C. D., & Vandaele, A. C. 2018, SSRv, 214, 10
Marcq, E., Salvador, A., Massol, H., & Davaille, A. 2017, JGRE, 122, 1539
Marrero, T. R., & Mason, E. A. 1972, JPCRD, 1, 3
McKenzie, D., Ford, P. G., Johnson, C., et al. 1992, JGR, 97, 13533
Mills, F. P. 1999, JGR, 104, 30757
Mogul, R., Limaye, S. S., Way, M. J., & Cordova, J. A. 2021, GeoRL, 48, e91327
Namiki, N., & Solomon, S. C. 1998, JGR, 103, 3655
Nikolaou, A., Katyal, N., Tosi, N., et al. 2019, ApJ, 875, 11
Nimmo, F. 2002, Geo, 30, 987
Nimmo, F. 2015, in Treatise on Geophysics, ed. G. Schubert, Vol. 9 (2nd ed.; Amsterdam: Elsevier), 201
Nimmo, F., & McKenzie, D. 1998, AREPS, 26, 23
Noack, L., Breuer, D., & Spohn, T. 2012, Icar, 217, 484
O'Rourke, J. G., & Korenaga, J. 2012, Icar, 221, 1043
O'Rourke, J. G., & Korenaga, J. 2015, Icar, 260, 128
O'Rourke, J. G., Wolf, A. S., & Ehlmann, B. L. 2014, GeoRL, 41, 8252
Odert, P., Lammer, H., Erkaev, N. V., et al. 2018, Icar, 307, 327
Ortenzi, G., Noack, L., Sohl, F., et al. 2020, NatSR, 10, 10907

Ostberg, C., & Kane, S. R. 2019, AJ, 158, 195
Pätzold, M., Häusler, B., Bird, M. K., et al. 2007, Natur, 450, 657
Persson, M., Futaana, Y., Ramstad, R., et al. 2020, JGRE, 125, e06336
Phillips, R. J., Bullock, M. A., & Hauck, S. A., II 2001, GeoRL, 28, 1779
Pluriel, W., Marcq, E., & Turbet, M. 2019, Icar, 317, 583
Pollack, J. B., & Black, D. C. 1982, Icar, 51, 169
Prather, M. J., & McElroy, M. B. 1983, Sci, 220, 410
Ramirez, R. M., Kopparapu, R. K., Lindner, V., & Kasting, J. F. 2014, AsBio, 14, 714
Raymond, S. N., Quinn, T., & Lunine, J. I. 2006, Icar, 183, 265
Raymond, S. N., Quinn, T., Lunine, J. I., et al. 2007, AsBio, 7, 66
Salvador, A., Massol, H., Davaille, A., et al. 2017, JGRE, 122, 1458
Schaber, G., Strom, R. G., Moore, H. J., et al. 1992, JGR, 97, 13257
Schaefer, L., & Sasselov, D. 2015, ApJ, 801, 40
Schaefer, L., Wordsworth, R. D., Berta-Thompson, Z., & Sasselov, D. 2016, ApJ, 829, 63
Schofield, N., Alsop, I., Warren, J., et al. 2014, Geo, 42, 599
Simons, M., Solomon, S. C., & Hager, B. H. 1997, GeoJI, 131, 24
Smrekar, S. E. 1994, Icar, 112, 2
Smrekar, S., Dyar, D, Helbert, J., et al. 2020, EPSC, 14, EPSC2020-447
Snellen, I., Guzman-Ramirez, L., Hogerheijde, M. R., et al. 2020, A&A, 644, L2
Sossi, P. A., Burnham, A. D., Badro, J., et al. 2020, SciA, 6, eabd1387
Stamnes, K., Tsay, S.-Chee, Jayaweera, K., & Wiscombe, W. 1988, ApOpt, 27, 2502
Strom, R. G., Schaber, G. G., & Dawson, D. D. 1994, JGR, 99, 10899
Stüeken, E. E., Kipp, M., Koehler, M. C., & Buick, R. 2016, ESRv, 160, 220
Tu, L., Johnstone, C. P., Güdel, M., & Lammer, H. 2015, A&A, 577, L3
Turcotte, D. L., & Schubert, G. 2002, Geodynamics (Cambridge: Cambridge Univ. Press)
Villanueva, G., Cordiner, M., Irwin, P., et al. 2021, NatAx, 5, 631
Warren, A., & Kite, E. 2021, BAAS, 53, 0101
Warren, J. K. 2010, ESRv, 98, 217
Way, M. J., & Del Genio, A. D. 2020, JGRE, 125, e06276
Way, M. J., Del Genio, A. D., Kiang, N. Y., et al. 2016, GeoRL, 43, 8376
Widemann, T., Ghail, R., Wilson, C. F., & Titov, D. V. 2020, AGUFM, P022-02
Wogan, N., Krissansen-Totton, J., & Catling, D. C. 2020, PSJ, 1, 58
Wordsworth, R. 2016, E&PSL, 447, 103
Wordsworth, R., & Pierrehumbert, R. 2014, ApJL, 785, L20
Wordsworth, R. D., & Pierrehumbert, R. T. 2013, ApJ, 778, 154
Wroblewski, F. B., Treiman, A. H., Bhiravarasu, S., & Gregg, T. K. P. 2019, JGRE, 124, 2233
Zahnle, K. J., Gacesa, M., & Catling, D. C. 2019, GeCoA, 244, 56
Zahnle, K. J., & Kasting, J. F. 1986, Icar, 68, 462
Zahnle, K. J., Kasting, J. F., & Pollack, J. B. 1988, Icar, 74, 62
Zahnle, K. J., Lupu, R., Catling, D. C., & Wogan, N. 2020, PSJ, 1, 11
Zeebe, R. E., & Westbroek, P. 2003, GGG, 4, 1104
Zolotov, M. 2019, in Oxford Research Encyclopedia of Planetary Science, ed. P. Read et al. (Oxford : Oxford Univ. Press),146
Zolotov, M. Y., Fegley, B., Lodders, K., et al. 1997, Icar, 130, 475